\documentclass[journal,12pt,onecolumn,draftclsnofoot,]{IEEEtran}
\usepackage[english]{babel}
\usepackage[dvipsnames,svgnames,x11names]{xcolor}
\usepackage[markup=underlined]{changes}

\IEEEoverridecommandlockouts

\ifCLASSINFOpdf

\else

\fi

\let\ssstyle=\scriptscriptstyle
\newcommand{\define}{\newcommand}
\newcommand{\bt}{\begin{tabular}}
\newcommand{\et}{\end{tabular}}
\newcommand{\be}{\begin{equation}}
\newcommand{\ee}{\end{equation}}

\newcommand{\bd}{\begin{displaymath}}
\newcommand{\ed}{\end{displaymath}}
\newcommand{\ba}{\begin{array}}
\newcommand{\ea}{\end{array}}
\newcommand{\bn}{\begin{enumerate}}
\newcommand{\en}{\end{enumerate}}
\newcommand{\bds}{\begin{description}}
\newcommand{\eds}{\end{description}}
\newcommand{\bi}{\begin{itemize}}
\newcommand{\ei}{\end{itemize}}
\newcommand{\bc}{\begin{center}}
\newcommand{\ec}{\end{center}}
\newcommand{\bqa}{\begin{eqnarray}}
\newcommand{\eqa}{\end{eqnarray}}

\define{\ul}{\noindent \underline}
\define{\Gamb}{{\bf \Gamma}}
\define{\Sigb}{{\bf \Sigma}}
\define{\Lambs}{{\bf \Lambda}_{s}}
\define{\LambR}{{\bf \Lambda}_{R}}
\define{\OmB}{{\bf \Omega}_{oB}}
\define{\Omo}{{\bf \Omega}_o}
\define{\Omb}{{\bf \Omega}}
\define{\Oms}{{\bf \Omega}_s}
\define{\alpb}{{\bf \alpha}}
\define{\alpbb}{\bar{\bf \alpha}}
\define{\betb}{{\bf \beta}}
\define{\betbb}{\bar{\bf \beta}}
\define{\tr}{\mbox{Tr}}
\define{\CR}{Cram\'{e}r-Rao~}
\define{\Mu}{{\it MUSIC}}
\define{\ES}{{\it ESPRIT}}
\define{\Mi}{{\it Min-Norm}}
\define{\SR}{{\it Subspace Rotation}}
\define{\ath}{{{\bf a}(\theta_k)}}
\define{\Ath}{{\bf A}(\theta)}
\define{\AthH}{{\bf A}^{H}(\theta)}
\define{\athd}{{{\bf a}^{(1)}(\theta_k)}}
\define{\bo}{{\bf 1}}
\define{\bz}{{\bf 0}}
\define{\ab}{{\bf a}}
\define{\bb}{{\bf b}}
\define{\cb}{{\bf c}}
\define{\cbb}{\bar{\bf c}}
\define{\db}{{\bf d}}
\define{\eb}{{\bf e}}
\define{\ebb}{\bar{\bf e}}
\define{\ebk}{{\bf e}_k}
\define{\ebbk}{\bar{\bf e}_k}
\define{\fb}{{\bf f}}
\define{\gb}{{\bf g}}
\define{\hb}{{\bf h}}
\define{\mb}{{\bf m}}
\define{\nb}{{\bf n}}
\define{\ob}{{\bf o}}
\define{\pb}{{\bf p}}
\define{\qb}{{\bf q}}
\define{\rb}{{\bf r}}
\define{\sbb}{{\bf s}}
\define{\tb}{{\bf t}}
\define{\ub}{{\bf u}}
\define{\vb}{{\bf v}}
\define{\wb}{{\bf w}}
\define{\st}{{\bf s}}
\define{\xb}{{\bf x}}
\define{\yb}{{\bf y}}
\define{\zb}{{\bf z}}
\define{\Ab}{{\bf A}}
\define{\Psib}{{\bf \Psi}}
\define{\Xib}{{\bf \Xi}}

\define\sT{{\ssstyle T}}
\define\sH{{\ssstyle H}}

\define{\Au}{{\Ab^{\uparrow}}}
\define{\Af}{{\bf A}_F}
\define{\As}{{\bf A}_S}
\define{\Bb}{{\bf B}}
\define{\Cb}{{\bf C}}
\define{\Cbh}{\hat{\bf C}}
\define{\Db}{{\bf D}}
\define{\Eb}{{\bf E}}
\define{\Fb}{{\bf F}}
\define{\Gb}{{\bf G}}
\define{\Hb}{{\bf H}}
\define{\Ib}{{\bf I}}
\define{\If}{{\bf I}_F}
\define{\Is}{{\bf I}_S}
\define{\Ibd}{{\bf I}^{\downarrow}}
\define{\Ibu}{{\bf I}^{\uparrow}}
\define{\Jb}{{\bf J}}
\define{\Kb}{{\bf K}}
\define{\Lb}{{\bf L}}
\define{\Lbb}{\bar{\bf L}}
\define{\Mb}{{\bf M}}
\define{\Nb}{{\bf N}}
\define{\Ob}{{\bf O}}
\define{\Pb}{{\bf P}}
\define{\Qb}{{\bf Q}}
\define{\Rb}{{\bf R}}
\define{\Rbh}{\hat{\bf R}}
\define{\Rsd}{\Delta {\bf R}_s}
\define{\Rba}{{\bf R}_A}
\define{\Rbse}{\stackrel{\sim}{{\bf R}_{s}}}
\define{\Rbb}{\bar{\bf R}}
\define{\Sb}{{\bf S}}
\define{\Tb}{{\bf T}}
\define{\Ub}{{\bf U}}
\define{\Vb}{{\bf V}}
\define{\Vn}{{\bf V_n}}
\define{\Wb}{{\bf W}}
\define{\Xb}{{\bf X}}
\define{\Yb}{{\bf Y}}
\define{\Zb}{{\bf Z}}
\define{\thb}{{\bf \Theta}}
\define{\Yd}{\Delta {\bf Y}}
\define{\Yt}{\tilde{\bf Y}}
\define{\Ut}{\tilde{\bf U}}
\define{\Vt}{\tilde{\bf V}}
\define{\Unt}{\tilde{\bf U}_o}
\define{\unt}{\tilde{\bf u}_o}
\define{\Ust}{\tilde{\bf U}_s}
\define{\ust}{\tilde{\bf u}_s}
\define{\Wt}{\tilde{\bf W}}
\define{\Wd}{\Delta \bf {W}}
\define{\Phb}{{\bf {\Phi}}}
\define{\Omegab}{{\bf {\Omega}}}
\define{\Gammab}{{\bf {\Gamma}}}
\define{\Phib}{{\bf {\Phi}}}
\define{\Phd}{\Delta \bf {\Phi}}
\define{\Phl}{{{\bf \Phi}^{\downarrow}}}
\define{\Phu}{{{\bf \Phi}^{\uparrow}}}
\define{\Psd}{\Delta \bf {\Psi}}
\define{\rt}{\tilde{r}}
\define{\rd}{\Delta r}
\define{\td}{\Delta \theta}
\define{\tl}{\tilde{\theta}}
\define{\ut}{\tilde{\bf u}}
\define{\Und}{\Delta {\bf U}_o}
\define{\Usd}{\Delta {\bf U}_s}
\define{\Us}{{\bf U}_s}
\define{\us}{{\bf u}_s}
\define{\Uo}{{\bf U}_o}
\define{\uo}{{\bf u}_o}
\define{\Vs}{{\bf V}_s}
\define{\vs}{{\bf v}_s}
\define{\Vo}{{\bf V}_o}
\define{\vo}{{\bf v}_o}
\define{\Aa}{||{\bf \alpha}_k||^{2}}
\define{\Usx}{{\bf U}_{sx}}
\define{\Usy}{{\bf U}_{sy}}
\define{\Usxd}{\Delta {\bf U}_{sx}}
\define{\Usyd}{\Delta {\bf U}_{sy}}
\define{\Usxt}{\tilde{\bf U}_{sx}}
\define{\Usyt}{\tilde{\bf U}_{sy}}
\define{\Usz}{{\bf U}_{sz}}
\define{\Uszd}{\Delta {\bf U}_{sz}}
\define{\Uszt}{\tilde{\bf U}_{sz}}
\define{\Ft}{\tilde{\bf F}}
\define{\Fd}{\Delta {\bf F}}
\define{\At}{\tilde{\bf A}}
\define{\Bt}{\tilde{\bf B}}
\define{\Bd}{\Delta {\bf B}}
\define{\ld}{\Delta \bar{\lambda}_k}
\define{\ldd}{\Delta \lambda_k}
\define{\lt}{\tilde{\lambda}}
\define{\lb}{\bar{\lambda}}
\define{\df}{\stackrel{\rm def}{=}}
\define{\diag}{\rm diag}
\define{\prf}{\noindent \underline{Proof:}\\}
\define{\pe}{\hfill $\Box$}
\define{\doi}{\stackrel{j\neq i}{j=1}}
\define{\dok}{\stackrel{j\neq k}{j=1}}
\define{\ds}{\displaystyle}
\define{\Usu}{{\bf U}_s^{\uparrow}}
\define{\Usl}{{\bf U}_s^{\downarrow}}
\define{\Unu}{{\bf U}_o^{\uparrow}}
\define{\Unl}{{\bf U}_o^{\downarrow}}
\define{\Oo}{{\bf \Omega}_{o}}
\define{\Ot}{{\tilde{\bf \Omega}}_{o}}
\define{\Os}{{\bf \Omega}_{s}}
\define{\Ou}{{\bf O}^{\uparrow}}
\define{\Od}{{\bf O}^{\downarrow}}
\define{\thd}{\Delta \theta}
\define{\Iu}{{\bf I}^{\uparrow}}
\define{\Il}{{\bf I}^{\downarrow}}
\define{\Otd}{{\tilde{\bf O}}^{\downarrow}}
\define{\Sr}{{\bf \Sigma}_s^{{1}\over{2}}}
\define{\Sp}{{\bf \Sigma}_s^{-1}}
\define{\Ss}{{\bf \Sigma}_s^{-{{1}\over{2}}}}
\define{\nun}{\underline{\bf n}}
\define{\nh}{\bar{\bf n}}
\define{\eq}[1]{(\ref{#1})}
\define{\dB}{\delta {\bf B}}
\define{\deb}{\delta b}
\define{\bul}{\underline{\beta}}
\define{\ape}{\stackrel{\cdot}{=}}

\define{\im}{\,\mbox{Im}}
\define{\re}{\,\mbox{Re}}

\define{\rmE}{\mathrm{E}}
\define{\rme}{\mathrm{e}}
\newcommand{\Expect}[1]{\mathop{\mathbb{E}}\left[ #1 \right ]}

\let\ssstyle=\scriptscriptstyle

\def\s0{{\ssstyle 0}}

\def\sT{{\ssstyle T}}
\def\sH{{\ssstyle H}}

\define{\Pib}{{\bf \Pi}}
\define{\bs}{\begin{slide}}
\define{\es}{\end{slide}}
\define{\xz}{\epsilon_{l,0}(n)}
\define{\xx}{\epsilon_{l,1}(n)}
\define{\xc}{\epsilon_{l,2}(n)}

\usepackage{amssymb,amsmath,epsfig}

\usepackage{color}
\usepackage{graphicx}
\usepackage{mathtools}

\usepackage{acronym}

 \usepackage[usenames,dvipsnames]{pstricks}
 \usepackage{epsfig}
 \usepackage{pst-grad} 
 \usepackage{pst-plot} 
 \usepackage[space]{grffile} 
 \usepackage{etoolbox} 
\usepackage{epstopdf}
\usepackage{cite}
\usepackage{mathtools}
\usepackage[algoruled,resetcount,linesnumbered]{algorithm2e}
\usepackage{float}
\newtheorem{proposition}{Remark}
\begin{document}
%
\title{Joint Frame Synchronization and Channel Estimation: Sparse Recovery Approach and USRP Implementation}


\author{ \IEEEauthorblockN{
\"{O}zg\"{u}r \"{O}zdemir\IEEEauthorrefmark{1}, Chethan Kumar Anjinappa\IEEEauthorrefmark{1}, Ridha
Hamila\IEEEauthorrefmark{2}, Naofal
Al-Dhahir\IEEEauthorrefmark{3}, and \.{I}smail~G\"{u}ven\c{c}\IEEEauthorrefmark{1}}
\\
\IEEEauthorblockA{\IEEEauthorrefmark{1}Department of Electrical \& Computer Engineering, North Caralina State University, Raleigh, NC\\
\IEEEauthorblockA{\IEEEauthorrefmark{2}Department of Electrical Engineering, Qatar University, Doha, Qatar}\\
\IEEEauthorblockA{\IEEEauthorrefmark{3} Electrical Engineering
Department, The University of Texas at Dallas, Richardson, TX \\Email:
\{oozdemi, canjina, iguvenc\}@ncsu.edu, hamila@qu.edu.qa, aldhahir@utdallas.edu}}\thanks{The work of R. Hamila and  N.Al-Dhahir was made possible by NPRP grant \# NPRP 8-627-2-260 from the Qatar National Research Fund (a member of Qatar Foundation). The statements made herein are solely the responsibility of the authors.  The work at NCSU was supported in part by NSF ACI-1541108.}}


\maketitle
\vspace{-1.5cm}
\begin{abstract}
Correlation-based techniques used for frame synchronization can suffer significant performance degradation over multi-path frequency-selective channels. In this paper, we propose a joint frame synchronization and channel estimation (JFSCE) framework as a remedy to this problem. This framework, however, increases the size of the resulting combined channel vector which should capture both the channel impulse response (CIR) vector and the frame boundary offset and, therefore, its estimation becomes more challenging. On the other hand, because the combined channel vector is sparse, sparse channel estimation methods can be applied. We propose several JFSCE methods using popular sparse signal recovery (SSR) algorithms which exploit the sparsity of the combined channel vector. Subsequently, the sparse channel vector estimate is used to design a sparse equalizer. Our simulation results and experimental measurements using software defined radios (SDRs) show that in some scenarios our proposed method improves the overall system performance significantly, in terms of the mean square error (MSE) between the transmitted and the equalized symbols compared to the conventional method.
\end{abstract}

\IEEEpeerreviewmaketitle

\section{Introduction}
In digital communication systems, the information symbols are usually transmitted within frames and to avoid degradation in performance, it is essential to determine the frame boundary correctly. Conventionally frame synchronization is performed by first correlating the received data sequence with a known training sequence and then determining time sample which gives the highest correlation as the frame boundary~\cite{massey&tcom:72}. However, in multi-path frequency-selective fading environments it is more difficult to identify the exact frame boundary as the delay spread becomes larger than the symbol duration (e.g., for vehicular channels with long delay spreads~\cite{6805659,6550867,V2X_ICC,V2X_VTC}) and the correlation peak gets widened. In addition to this, if the location of the strongest multi-path component happens to be somewhere other than the first tap, the dominant component location will give the highest correlation resulting in an incorrect frame boundary selection.

The multi-path channel can be converted into a single-tap flat-fading channel by using equalization techniques and this can improve the frame synchronization accuracy. However, the channel estimate required for equalization is often calculated using the training symbols assuming perfect knowledge of the frame boundary itself. Therefore, JFSCE can be an effective technique to improve equalization performance.

We tackle this problem by treating the frame boundary offset as an unknown delay introduced to the channel which forms the combined channel. The combined channel estimation provides channel estimate as well as frame synchronization. The equalizer is designed based on the combined channel estimate. Obviously, introducing the delay into the channel, increases the length of the channel impulse response (CIR) by an amount as large as the duration of the frame length and this increases the computational complexity of the channel estimation. However, the combined channel is a sparse vector because introducing delay into the channel is equivalent to padding the CIR with zeros. Therefore, the number of non-zero CIR taps that needs to be estimated will be the same. As a result, complexity can be reduced by using sparse channel estimation methods.

Sparse channel estimation has been investigated in~\cite{cotter&tc:2002} using a matching pursuit (MP) algorithm~\cite{mallat@tsp:93,cotter&visp:99} while a method based on least mean squares (LMS) is proposed in~\cite{gui&wcnc:13}. In~\cite{karabulut&vtc:2004} the orthogonal matching pursuit (OMP) algorithm instead of MP is used for sparse channel estimation. None of the methods in~\cite{cotter&tc:2002,mallat@tsp:93,cotter&visp:99,gui&wcnc:13,karabulut&vtc:2004}  considers frame synchronization while addressing sparse channel estimation. There exists earlier work on JFSCE. However, none of them exploits the sparsity of the combined channel vector. JFSCE is studied for OFDM systems in~\cite{wen@wicom:07}, for CDMA systems in \cite{marey@vtc:07}, and for optical communication systems in~\cite{cheng&spie:16}. To the best of our knowledge, JFSCE using sparse recovery algorithms has not been studied in the literature.

In this paper, we propose a JFSCE method which uses some of the existing sparse signal recovery (SSR) algorithms to exploit the sparsity of the combined CIR vector to obtain an estimate of the combined CIR vector. In particular, optimization based, greedy, thresholding based, and Bayesian methods are utilized, and their performance and complexity trade-offs are compared in the context of the proposed JFSCE framework. The focus of this work is not the development of these SSR algorithms but adopt the best algorithms from each class to a new problem. The combined channel estimate we get using the proposed JFSCE method is used to design a sparse equalizer that minimizes the mean square error (MSE) between the equalizer output and the transmitted symbols, to illustrate the performance of the proposed method. The performance improvement achieved by the proposed JFSCE method is demonstrated by both simulation results and experimental results using a universal software radio peripheral (USRP) testbed. We show that in situations where the conventional method fails to identify the correct frame boundary or the number of taps assumed in the channel estimate is less than the actual length of the channel, the proposed method improves the system performance significantly.

The rest of this paper is organized as follows.  Section~II
describes the system model. In Section III, the proposed JFSCE framework is described along with the conventional method and classical JFSCE method. In Section IV, we provide an overview of the sparse signal recovery algorithms that we use  within the JFSCE framework. An equalizer design based on the computed sparse channel estimate is discussed in Section V. Simulation results in Section VI and experimental USRP testbed results in Section VII demonstrate the performance improvements of our proposed method. The paper is concluded in Section VIII.

{\bf Notation}: Vectors and matrices are represented by lower-case and upper-case boldface letters, respectively. The transpose and conjugate transpose are denoted by  $(.)^{\sT}$ and $(.)^{\sH}$, respectively. $\Expect{.}$ is the expectation operator. $\diag(x_1,x_2,\dots,x_n)$ represents the diagonal matrix which maps an n-tuple to the corresponding diagonal matrix. For an integer $k \in \mathcal{Z}$, we
use the shorthand notation $[k]$ for the set of non-negative integers $\{1,2,\dots, k\}$. The support of a vector $\xb \in \mathcal{C}^N$ is the index set of non-zero entries of $\xb$, i.e., supp(\xb) = $\{j \in [N]: \xb_j \neq 0\}$. The vector $\xb$ is called $k$-sparse if at most $k$ of its entries are non-zero.

\begin{figure}[!t]
\centering
\includegraphics[scale=1.1]{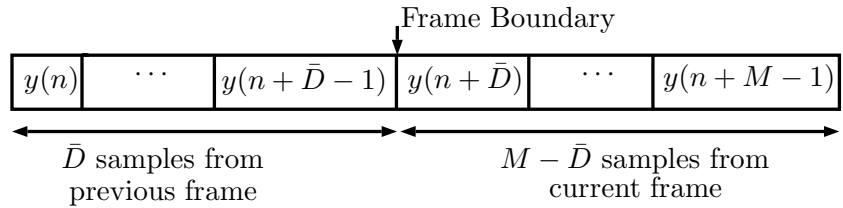}
\caption{The received samples and the illustration of frame boundary $\bar{D}$ for a data frame.}
\label{regframe:fig}
\end{figure}

\section{System Model}
A frame-based communication system over a multi-path frequency-selective channel is considered where the information symbols are transmitted in frames. In order to recover the transmitted symbols, the receiver needs to estimate the frame boundary. This is referred to as frame synchronization. Furthermore, channel estimation needs to be performed at the receiver. The channel estimate is used for channel equalization and demodulation to recover the information symbols at the receiver. We denote the symbol-spaced multi-path CIR by the vector $\hb=\left [ h_0, h_1, \cdots, h_L \right ]^{\sT}$ where $L$ is the CIR memory. Suppose that the frame synchronization is performed prior to channel estimation, then, assuming symbol-spaced sampling at the receiver, the $n$-th received symbol after frame synchronization can be expressed as follows
\begin{align}
  y(n)&=\sum_{l=0}^{L} x(n-l) h_l + z(n) \\
  &= x(n) h_0 + x(n-1) h_1 +\cdots + x(n-L) h_L +z(n),
\end{align}
where $x(n)$ is the transmitted symbol and $z(n)$ is the additive white Gaussian noise (AWGN) symbol at time $n$. Note that, here $x(n)$ is multiplied by $h_0$ to construct $y(n)$. 

\begin{proposition} We use \emph{symbol spaced CIR} and \emph{symbol spaced sampling} at the receiver for clarity of the presentation throughout the paper. However the channel estimation and equalization as discussed later in the paper can be performed assuming fractionally spaced methods~\cite{cioffi:fractional} as well. Fractionally spaced methods perform better than the symbol spaced methods at the price of increased computational complexity.
\end{proposition}

Let us now suppose frame synchronization is not performed prior to channel estimation. In that case, the received signal is given by
 \begin{align}
 \label{delayD:eq} y(n)&=\sum_{l=0}^{L} x(n-D-l) h_l + z(n),
\end{align}
where $D$ is the delay, in symbol periods, between the transmitter and the receiver. Note that, $h_0$ now multiplies $x(n-D)$ instead of $x(n)$. Suppose that the number of samples in the transmitted frames is $M$ and that we arbitrarily collect $M$ samples $\{y(n), \cdots, y(n+M-1) \}$ without knowledge of the frame boundary denoted by $\bar{D}$. The frame boundary, $\bar{D}$, is a random number between 0 and $M-1$ which depend on at which sample we start collecting the samples. If all of the $M$ received samples correspond to the same frame, this means frame boundary is equal to zero. Otherwise, some initial samples actually belong to the previous frame. This is illustrated in Fig.~\ref{regframe:fig}.
The delay $D$ in~(\ref{delayD:eq}) and the frame boundary $\bar{D}$ are related by
\be
\label{barD:eq}
D=mM+\bar{D},
\ee
where $m=0,1,\cdots$ is an integer. As shown in Fig.~\ref{regframe:fig} knowledge of $m$ is not required to divide the received samples into frames. Therefore, throughout the paper, we only investigate finding $\bar{D}$ for frame synchronization.

\begin{figure}[!t]
\centering
\includegraphics[scale=1.1]{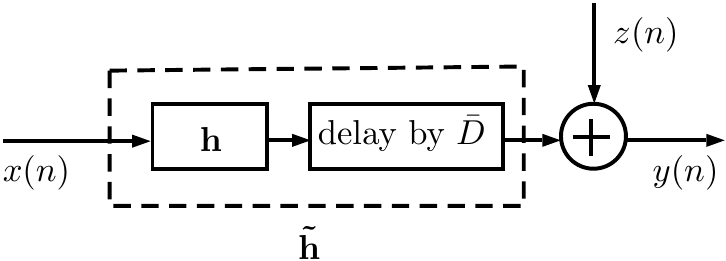}
\caption{The system model with combined channel, $\tilde{\hb}$, that takes the multi-path channel, $\hb$,  and delay into consideration.}
\label{sysmodel:fig}
\end{figure}

Although frame synchronization and channel estimation can be performed separately, we may treat them jointly as well, by defining the following delayed and zero-padded combined CIR vector $\tilde{\hb}$
\be
\label{htilde:eq}
\tilde{\hb}= [\underbrace{0, 0, \cdots, 0}_{\bar{D} \: \mathrm{zeros}} \underbrace{h_0, h_1, \cdots, h_L}_{\hb^{\sT}}, \underbrace{0, 0, \cdots, 0}_{M-\bar{D}-1 \: \mathrm{zeros}}  ]^{\sT}.
\ee
Fig.~\ref{sysmodel:fig} illustrates the system model with combined channel that takes the multi-path channel and delay into consideration.
In~(\ref{htilde:eq}), the $M-\bar{D}-1$ zeros at the end do not have any effect as far as the input and output of the system are concerned. However, they ensure that the length of $\tilde{\hb}$ remains fixed at $M+L$ which is independent of the value of $\bar{D}$. Note that, whether the original channel, $\hb$, is sparse or not, as long as $M\gg 1$, the combined CIR vector, $\tilde{\hb}$, will be a sparse vector.

\section{Proposed Joint Frame Synchronization and Channel Estimation Framework}
\label{joint:sec}

\begin{figure}[!t]
\centering
\includegraphics[scale=1]{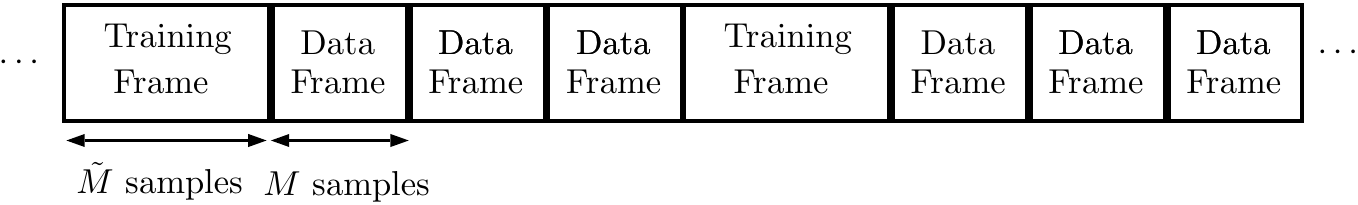}
\caption{The frame structure for the proposed JFSCE framework. Training frames are periodically transmitted every $P$ data frames. Here we assume $P=4$. The size of the training frame is $\tilde{M}$ samples and the size of the data frame is $M$ samples.}
\label{frame:fig}
\end{figure}

To perform JFSCE, we assume that a known training frame with size $\tilde{M} > M$ is periodically transmitted as shown in Fig.~\ref{frame:fig}.
It should be clear later in this section why the size of the training frame needs to be larger than the size of the data frame. The estimate of combined channel, $\tilde{\hb}$, is calculated using the training frame. This estimate is then used to design the equalizer that is used to demodulate the data frames until the next $\tilde{\hb}$ estimate is calculated. This occurs when the next training frame is transmitted again after $P$ frames. The period $P$ should be determined according to the channel coherence time. Keeping it small increases the overhead in the system. On the other hand, keeping it large degrades the channel estimation accuracy. The input-output relation of the system model illustrated in Fig.~\ref{sysmodel:fig} is
\be
\label{yn:eq}  y(n)=\sum_{l=0}^{M+L-1} x(n-l) \tilde{h}_l + z(n),
\ee
where $\tilde{h}_l$ is the $l$-th element of $\tilde{\hb}$.

When the current frame is a data frame, the receiver will collect $M$ samples as in Fig.~\ref{regframe:fig} where $M$ is the number of samples in a data frame. Similarly, when the current frame is a training frame, the receiver collects $\tilde{M}$ samples $\{y(n), \cdots, y(n+\tilde{M}-1) \}$ where $\tilde{M}$ is the number of samples in a training frame.
Note that, because $\bar{D}$ is an integer in $[0, M-1]$, in the extreme case, $\bar{D}=M-1$ and the samples $y(n+M)$, $y(n+M+1)$, $\cdots$, $y(n+\tilde{M}-1)$ are guaranteed to be in the training frame rather than in the frame preceding the training frame. In JFSCE, a subset of these samples, namely $y(n+\tilde{M}-N_{\mathrm{E}})$, $y(n+\tilde{M}-N_{\mathrm{E}}+1)$, $\cdots$, $y(n+\tilde{M}-1)$ are used where $N_{\mathrm{E}}$ is the number of equations. Using (\ref{yn:eq}) and the sample $y(n+\tilde{M}-1)$, the first equation is given by
\be
\label{first:eq}
y(n+\tilde{M}-1)=\sum_{l=0}^{M+L-1} x_{t,\tilde{M}-1-l} \tilde{h}_l + z(n+\tilde{M}-1),
\ee
where $x_{t,0},\cdots,x_{t,\tilde{M}-1}$ are the known transmitted symbols in the training frame. Similarly, the last equation is given by
\be
\label{eqnNE:eq} y(n+\tilde{M}-N_{\mathrm{E}})=\sum_{l=0}^{M+L-1} x_{t,\tilde{M}-N_{\mathrm{E}}-l} \tilde{h}_l + z(n+\tilde{M}-N_{\mathrm{E}}).
\ee

In~(\ref{eqnNE:eq}) when $l=M+L-1$, we have $x_{t,\tilde{M}-N_{\mathrm{E}}-l}=x_{t,\tilde{M}-N_{\mathrm{E}}-M-L+1}$. Therefore, to guarantee that $x_{t,\tilde{M}-N_{\mathrm{E}}-l}$ is a valid training symbol, the following relation needs to be satisfied:
\be
\tilde{M}-M-L-N_{\mathrm{E}}+1 \geq 0.
\ee
Therefore, $\tilde{M}\geq M+L+N_{\mathrm{E}}-1$. Choosing $\tilde{M}=M+L+N_{\mathrm{E}}-1$ (to reduce the training overhead) and stacking $N_{\mathrm{E}}$ received samples in a column vector we get
\be
\label{sys:eq}
\yb=\tilde{\Xb}_t \tilde{\hb} +\zb,
\ee
 where
 \be
\yb=\left [y(n+\tilde{M}-1), y(n+\tilde{M}-2), \cdots, y(n+\tilde{M}-N_{\mathrm{E}})   \right ] ^{\sT}
 \ee
 is the known received vector of size $N_{\mathrm{E}}$ and
 \be
 \tilde{\Xb}_t=\left [ \begin{array}{cccc}
                         x_{t,\tilde{M}-1} & x_{t,\tilde{M}-2} & \cdots & x_{t,N_{\mathrm{E}}-1} \\
                         x_{t,\tilde{M}-2} &  x_{t,\tilde{M}-3} &  & \vdots \\
                         \vdots &  & \ddots &  \\
                         x_{t,\tilde{M}-N_{\mathrm{E}}} & \cdots &  & x_{t,0}
                       \end{array}
 \right]
 \ee
is the $N_{\mathrm{E}} \times (M+L) $ measurement matrix constructed from known training symbols. Finally, $\zb$ is the noise vector
 \be
  \zb\!=\!\!\left [z(n+\tilde{M}-1), z(n+\tilde{M}-2), \cdots, z(n+\tilde{M}-N_{\mathrm{E}})   \right ] ^{\sT}.
 \ee

\subsection{Conventional Method}
\label{conventional:sec}
In the conventional method, frame synchronization and channel estimation are  performed  separately. First, the received data symbols are cross-correlated with the training symbols to obtain the estimate of the frame boundary, $\hat{\bar{D}}$ in $[0,M-1]$\footnote{Although the cross-correlation method is optimal for frequency-flat channels, it becomes suboptimal for multi-path frequency-selective channels.}. Then, (\ref{sys:eq}) can be expressed compactly as follows
\begin{equation}\label{Conv_Eq}
\yb = {\Xb}_t {\hb} +\zb,
\end{equation}
where
 \be
  {\Xb}_t=\left [ \begin{array}{cccc}
                         x_{t,\tilde{M}-\hat{\bar{D}}-1} & x_{t,\tilde{M}-\hat{\bar{D}}-2} & \cdots & x_{t,\tilde{M}-\hat{\bar{D}}-L-1} \\
                         x_{t,\tilde{M}-\hat{\bar{D}}-2} &  x_{t,\tilde{M}-\hat{\bar{D}}-3} &  & \vdots \\
                         \vdots &  & \ddots &  \\
                         x_{t,\tilde{M}-\hat{\bar{D}}-N_{\mathrm{E}}} & \cdots &  & x_{t,\tilde{M}-\hat{\bar{D}}-L-N_E}
                       \end{array}
 \right]
 \ee
is the $N_E \times (L+1)$ measurement matrix constructed from known training symbols.

To gain some insight from (\ref{Conv_Eq}), the following observations are in order:
\begin{enumerate}
\item The first equation which includes the first row of ${\Xb}_t$ is given in (\ref{eqnNE:eq}) where $\bar{D}$ is replaced by $\hat{\bar{D}}$.
\item ${\Xb}_t$ is a sub-matrix of $\tilde{\Xb}_t$. For example if $\tilde{\bar{D}}=0$, then ${\Xb}_t$ consists of the first $L+1$ columns of $\tilde{\Xb}_t$. If $\hat{\bar{D}} =  M - 1$, then ${\Xb}_t$ consists of the last $L+1$ columns of ${\Xb}_t$.
\item $\hb$ may or may not be a sparse vector. On the other hand, $\tilde{\hb}$ will always be a sparse vector, .
\end{enumerate}
If $\hb$ is not sparse, then the classical least-squares solution $\hat{\hb} = {\Xb}_t^{\dagger} \yb$, described in the next sub-section, is a reasonable low-complexity approach to solve this problem although sparse methods can still be applied if $\hb$ is a sparse CIR vector. Once $\hat{\hb}$ is obtained, the combined CIR vector estimate becomes
\begin{equation}
\hat{\tilde{{\hb}}}_{conv}= [\underbrace{0, 0, \dots, 0}_\text{$\tilde{\bar{D}}$ zeros}, \tilde{\hb}^T, \underbrace{0, 0, \dots, 0}_\text{$M-\tilde{\bar{D}} -1$ zeros}]^T.
\end{equation}
When using the conventional method, the error in determining the frame boundary may lead to inaccuracies in the channel estimate.

\subsection{Classical JFSCE Method}
 \label{classical:sec}
In order to recover $\tilde{\hb}$, classical JFSCE method  minimizes the least-square (LS) of the error vector, $\eb = \yb - \tilde{\Xb}_t \tilde{\hb}$, which leads to the classical solution given by
\be
\label{jointclassical:eq}
\hat{\tilde{\hb}}_{\mathrm{classical}}=\tilde{\Xb}_t^{\dagger}\yb,
\ee
where $\tilde{\Xb}_t^{\dagger}$ is the pseudo-inverse of $\tilde{\Xb}_t$ and this solution is called the minimum-norm solution if $\tilde{\Xb}_t$ is a wide matrix. Two major problems with the classical solution are as follows:
\begin{enumerate}
\item To obtain an accurate estimate of $\tilde{\hb}$, the number of equations $N_{\mathrm{E}}$ may be prohibitively large which increases the required number of measurements and computational complexity.
\item The solution is not guaranteed to be a sparse solution although we know that $\tilde{\hb}$ is a sparse vector.
\end{enumerate}
When the channel vector is sparse, the problem to recover $\tilde{\hb}$ from $\yb$ becomes a sparse recovery problem and the sparse signal recovery (SSR) algorithms become an effectual means to recover the sparse vector~$\hat{\tilde{\hb}}$. Sparse channel vectors can be acquired by the SSR-based JFSCE method using fewer measurements than what the classical JFSCE method requires. In Section \ref{SSR_Algorithms}, we discuss some of the popular SSR algorithms that recover the original sparse channel vector within our JFSCE framework in an under-determined setting, which results in a relatively lower training overhead and computational cost compared to the classical JFSCE method. 

\section{Sparse Signal Recovery Based JFSCE Method} 
\label{SSR_Algorithms}
In this section, we adapt some of the popular sparse signal recovery (SSR) algorithms for JFSCE to recover the combined CIR sparse vector from an under-determined set of measurements. We note that there are plethora of algorithms in the literature and one can adapt different SSR algorithms based on different criteria such as the recovery performance and the computational complexity. Broadly speaking, SSR algorithms can be grouped into 4 categories: optimization algorithms, greedy algorithms, Bayesian algorithms, and thresholding algorithms. In this paper, we consider the following four algorithms to be used in the JFSCE framework and compare their performance and complexity trade-offs:  
\begin{itemize}
\item Optimization: Re-weighted $\ell_1$ (R-$\ell_1$)-based JFSCE method
\item Greedy: OMP and CoSaMP-based JFSCE method
\item Bayesian: SBL based JFSCE method
\item Thresholding: EMGMAMP-based JFSCE method
\end{itemize}
The rationale behind the choice of these SSR algorithms is that they represent different points on the performance-complexity tradeoff curve.
In the subsequent sub-sections, we briefly discuss each of the above mentioned SSR-based JFSCE methods. For detailed analysis of the SSR algorithms, readers may refer to the corresponding references.

\subsection{R-$\ell_1$-based JFSCE Method}
Re-weighted $\ell_1$ algorithm (R-$\ell_1$)~\cite{Rl1} is categorized under the optimization methods, where the underlying principle to recover the sparse vector is to solve an optimization problem. Ideally, the sparse channel vector can be obtained by solving 
\begin{equation} \label{l0_norm}
\arg\underset{\tilde{\hb} \in \mathbb {C}^{M+L}}{\min} \Vert \tilde{\hb}\Vert_0,~\mathrm{s.t.}\: \yb = \tilde{\Xb}_t \tilde{\hb},
\end{equation}
where $\Vert \tilde{\hb}\Vert_0$ is the $\ell_0$-norm of $\tilde{\hb}$\footnote{$\ell_0$-norm of a vector is the number of non-zero elements in that vector.}. Taking the measurement error into account, the problem in~(\ref{l0_norm}) can be formulated as follows
\be
\label{sparse:eq}
\arg\underset{\tilde{\hb} \in \mathbb {C}^{M+L}}{\min}  \Vert\yb- \tilde{\Xb}_t \tilde{\hb}\Vert_2^2 + \lambda \Vert \tilde{\hb}\Vert_0,
\ee
where $\lambda$ is a positive penalty parameter that strikes a trade-off between the sparsity and measurement-fidelity of the solution. Unfortunately, the $\ell_0$-norm is non-convex and requires combinatorial complexity to find the solution. The most popular approach to overcome the computational complexity issue is to rely on iterative re-weighting schemes which produce weights as the optimization progresses \cite{Rl1,Rl2,Rl1_l2_Wipf}. One such variant is R-$\ell_1$, where at $k^{th}$ iteration we solve 
\be
\label{sparse:eq}
(x^{k+1}) \rightarrow \arg\underset{\tilde{\hb} \in \mathbb {C}^{M+L}}{\min}  \Vert\yb- \tilde{\Xb}_t \tilde{\hb}\Vert_2^2 + \lambda \sum_{i} w_{i}^k\vert \hb_i \vert,
\ee
where the term $w_{i}^k = (\vert \hb_i^k \vert + \epsilon)^{-1}$ is the $\ell_1$ weighting factor that needs to be updated after each iteration. The regularization parameter $\lambda$ was chosen as $4 \sigma \sqrt{M + L - k}$, rule of thumb which is motivated by a theoretical analysis \cite{Lambda}, where, $\sigma^2$ is the variance of the \textit{i.i.d} AWGN noise $z$, and $k$ is the sparsity level. The solution to the problem (\ref{l0_norm}) can also be linked to the output of basis pursuit denoising (BPDN) algorithm \cite{BPDN} and the least absolute shrinkage and selection operator (LASSO) algorithm \cite{LASSO}. The existence of efficient solvers of the above variants makes them excellent choices. However, the computational complexities of these methods are higher and it is challenging to implement them in practice.

\subsection{OMP and CoSaMP-based JFSCE Method}
Orthogonal matching pursuit (OMP)~\cite{OMP} and compressive sampling matching pursuit (CoSaMP)~\cite{cosamp} are among the popular greedy algorithms. The algorithms under this category often produce results that are comparable to the optimization-based algorithms at a relatively lower computational complexity. The underlying principle is to find the support set in an iterative manner based on the greedy (projection) strategy and then solve a LS optimization that best fits the collected measurements in the subspace spanned by all previously selected columns. The projection part is the most costly in terms of computational complexity. 

For instance, OMP adds an index $j_{n+1}$ to a target support $\mathcal{S}^{n+1}$ at each iteration and updates the target channel vector $\hat{\tilde{\hb}}$ as the vector supported on the target support that best fits the collected measurement $\yb$ for the JFSCE problem. The algorithm is formally defined in Algorithm \ref{OMP_PseudoCode}.

\begin{algorithm}[h!]
    \SetKwInOut{Input}{Input}
    \SetKwInOut{Output}{Output}
    \Input{$\tilde{\Xb}_t$, $\yb$}
    \Output{$\hat{\tilde{\hb}}$}
    \For{Until the stopping criterion is met}
      {
      $\mathcal{S}^{n+1} = \mathcal{S}^{n} \cup \{j_{n+1}\}$;\\
      $j_{n+1} = \text{arg} \underset{j \in [N]}{\max} \{|\tilde{\Xb}_t^*(\yb- \tilde{\Xb}_t \hat{\tilde{\hb}}^n)|_j\}$;\\
      $\hat{\tilde{\hb}}^{n+1}= \text{arg} \underset{z \in \mathcal{C}^{M+L}}{\min} \{||\yb - \tilde{\Xb}_t \tilde{\hb} ||_2, \text{supp($\tilde{\hb}$)} \subset \mathcal{S}^{n+1}\}$
      }
      {
        return $\hat{\tilde{\hb}}$;
      }
    \caption{Orthogonal Matching Pursuit}\label{OMP_PseudoCode}
\end{algorithm}

The main weakness of the OMP-based JFSCE method is that, once an incorrect index is added to the target support, it remains in all the subsequent target supports resulting in an erroneous support set. The recovery performance can be further improved by adopting variants of the OMP such as the generalized OMP algorithm~\cite{GOMP} and CoSaMP algorithm~\cite{cosamp} at an additional computational complexity.

\subsection{SBL-based JFSCE Method}
The above mentioned class of SSR algorithms do not take into account the structure of the sparse channel vector and noise, and hence, are not designed to exploit the known properties of the noise covariance matrix and/or any other structural properties of the sparse channel vector. Whereas, the Bayesian methods provide a natural and disciplined way of utilizing the apriori information of the sparse vector and elegantly incorporating the noise covariance matrix structure and the correlation constraints into the SSR problem. 

In the Bayesian framework, a family of algorithms called Sparse Bayesian Learning (SBL) algorithms~\cite{SBL} has been developed to find robust solutions to the SSR problems. In this approach, the target sparse channel vector ${\hb}$ is assumed to be $\mathcal{N}(0,\Gamma)$, where $\Gamma = \diag(\gamma(1),\ldots,\gamma(N))$ represents the unknown hyper-parameters. The estimation
of ${{\hb}}$ reduces to estimation of the hyper-parameter vector $\Gamma$ which can be performed in SBL using the iterative EM framework. The expectation (E-step) in the $i$-th iteration evaluates the log-likelihood function $L(\Gamma|\Gamma(i))$ given by
\begin{equation}
L(\Gamma|\Gamma(i)) = \mathop{\mathbb{E}}_{\hat{\tilde{h}}|{\yb};\Gamma(i)} {\log p({\yb},\hat{\tilde{h}};\Gamma)}~.
\end{equation}
The maximization (M-step) yields the hyper-parameter vector estimate ~$\Gamma(i+1)$ by maximizing the log-likelihood function at the $i$-th iteration with respect to $\Gamma$ as follows:
\be
\Gamma(i+1) = \arg \underset{\Gamma}{\max}L(\Gamma |\Gamma(i)).
\ee
There are numerous variants of SBL algorithm like T-SBL algorithm, TMSBL algorithm~\cite{TSBL}, etc., which consider temporal correlation. In general, SBL-based JFSCE method performs well but its computational complexity is high. 

\subsection{EMGMAMP-based JFSCE Method} 
The expectation-maximization Gaussian mixture AMP (EMGMAMP)~ \cite{EMGMAMP} algorithm is a powerful thresholding-based algorithm. It combines two powerful inference frameworks, namely expectation maximization (EM) and approximate message passing (AMP). The EMGMAMP algorithm assumes a sparsity promoting $i.i.d.$ Gaussian mixture prior and an additive Gaussian noise prior for $\hb$ and $\zb$, respectively, which are given by:
\begin{eqnarray}
\begin{aligned}
&p_{\tilde{\hb}}(\tilde{\hb}) = \eta f_{\tilde{\hb}}(\tilde{\hb}) + (1-\eta) \delta_{\tilde{\hb}}\\
&p_{\zb}(\zb) = \mathcal{N}(\zb;0,\phi),
\end{aligned}
\end{eqnarray}
where $\eta$ is the sparsity rate and $\delta(\cdot)$ is the Dirac-delta function. The EMGMAMP assumes $f_{\tilde{\hb}}$ to be an $L$-term Gaussian mixture. Mathematically, $f_{\tilde{\hb}} = \sum_{l=1}^L w_l \mathcal{N}(\tilde{\hb};\mu_l,\gamma_l)$. where $w_l$ is the weight associated with the $l^{th}$ Gaussian mixture and satisfies $\sum_{l=1}^L w_l = 1$. The terms $\mu_l$ and $\gamma_l$ are the mean and variance associated with the $l^{th}$ Gaussian mixture component, respectively. The noise is assumed to be $i.i.d.$ Gaussian with zero mean and $\phi$ variance. The means and variances of the posterior $p(\hb|\yb)$ are evaluated using the Generalized Approximate Message Passing (GAMP) framework \cite{GAMP_Sundeep}. Upon running GAMP, the framework utilizes the EM algorithm to update the parameters $\eta,\wb,\boldsymbol{\mu},\boldsymbol{\gamma},$ and $\boldsymbol{\phi}$ until convergence. We skip further technical details on the EMGMAMP noting that we adapt EMGMAMP within the JFSCE framework; interested readers may refer to \cite{EMGMAMP,GAMP_Sundeep}. The key appealing features of the algorithms falling under this class are their generality and computational scalability. Thus, they are well suited for practical implementations of sparse recovery. 

In the remainder of this paper, we consider these SSR-based JFSCE methods and evaluate their performance and complexity trade-offs.


\section{Sparse Equalizer Design Based on Channel Estimates}
\label{equalizer:sec}
\begin{figure}[!t]
\centering
\centerline{\includegraphics[scale=0.95]{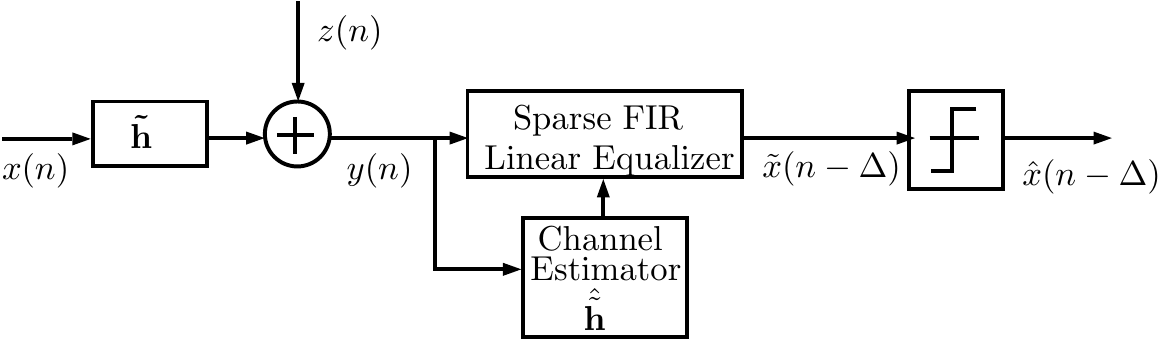}}
\caption{The complete system model illustrating the combined channel, channel estimator and the equalizer.}
\label{comsys:fig}
\end{figure}

In communication systems with multi-path frequency selective fading inter-symbol interference (ISI) degrades the system performance. To mitigate the ISI an equalizer should be implemented before decisions on the symbols are made. Fig.~\ref{comsys:fig} illustrates the complete system model with the combined channel as well as the equalizer before the thresholding device which makes decisions.   
The equalizer is designed using the channel estimates from the previous section. Therefore, the performance of the equalizer will depend on the accuracy of the particular channel estimate used. As a result, the performance of different channel estimators from the previous section can be compared by evaluating the equalizer performance. 

To reduce the complexity of the equalizer, we implement the sparse finite impulse response (FIR) linear equalizer design of~\cite{abbasi&globalsip:2015,abbasi&twc:17}. The output of the equalizer is the soft estimate of the transmitted symbol $x(n)$ and can be expressed as $\tilde{x}(n)=\sum_{k=0}^{N-1} y(n-k)w_k,$
where $w_k$ is the $k$-th element of the equalizer vector $\wb$ and $N$ is the length of the equalizer. The performance metric adopted is the MSE defined as follows
\be\label{Equalization_MSE}
\mathrm{MSE}=\Expect{|\tilde{x}(n)-x(n-\Delta)|^2},
\ee
where $\Delta$ is the equalizer delay which is optimized to reduce the MSE. Finally, the transmitted symbols' estimates are calculated using a decision device based on the type of signal constellation used.

Note that we prefer to use the MSE from the equalizer output instead of MSE of the channel estimates because our end-to-end metric is the MSE between transmitted and estimated symbols. A better channel estimate may sometimes result in a worse equalizer performance. As an example, assume that in a channel estimate the locations of the non-zero channel taps are estimated correctly  but this estimate has large channel estimation MSE. Another channel estimate has a better channel estimate but it also has additional small non-zero components that do not exist in the actual channel. These small coefficients do not affect the channel estimation MSE significantly. However, they can cause the equalizer performance of the second channel estimate to be worse than the equalizer performance of the first channel estimate. 

In the next section, computer simulations will be performed to assess the performance of the proposed SSR-based JFSCE methods. Then, Section VI presents the implementation of the OMP-based JFSCE method on a software defined radio testbed.

\section{Simulation Results}
\begin{figure}[t]
\centering
\centerline{\includegraphics[scale=0.6]{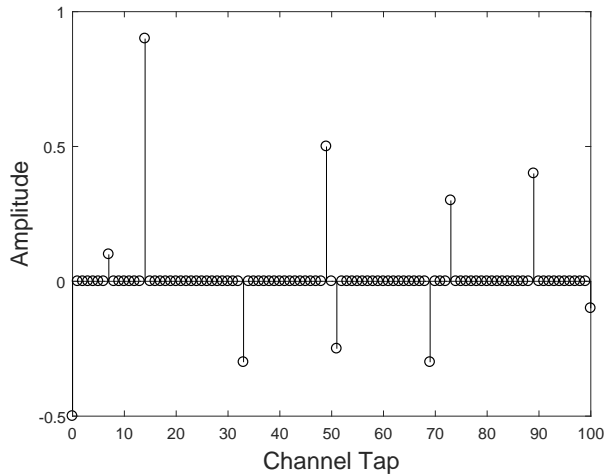}}
\caption{The CIR $\hb$ used in the simulations.}
\label{channel:fig}
\end{figure}
Monte Carlo simulation results are presented to evaluate the performance of the SSR-based JFSCE methods. The CIR that is shown in Fig.~\ref{channel:fig} is used in the simulations. This CIR is similar to that used in~\cite{cotter&tc:2002}. This CIR has $L+1=101$ taps with 10 non-zero taps given by: $h_0 = -0.5$, $h_7 = 0.1$, $h_{14} = 0.9$, $h_{33} = -0.3$, $h_{49} = 0.5$, $h_{51} = -0.25$, $h_{69} = -0.3$, $h_{73} = 0.3$, $h_{89} = 0.4$, and $h_{100} = -0.1$. For channel estimation, the data frame length is $M=1000$ and $N_{\mathrm{E}}=148$ equations are used. The length of the training frame is $\tilde{M}=M+L+N_{\mathrm{E}}-1=1247$. The frame boundary is arbitrarily set to $\bar{D}=500$.


\subsection{MSE Performance}
\label{simulationA:sec}
\begin{figure}[t!]
\centering
\centerline{\includegraphics[scale=0.8]{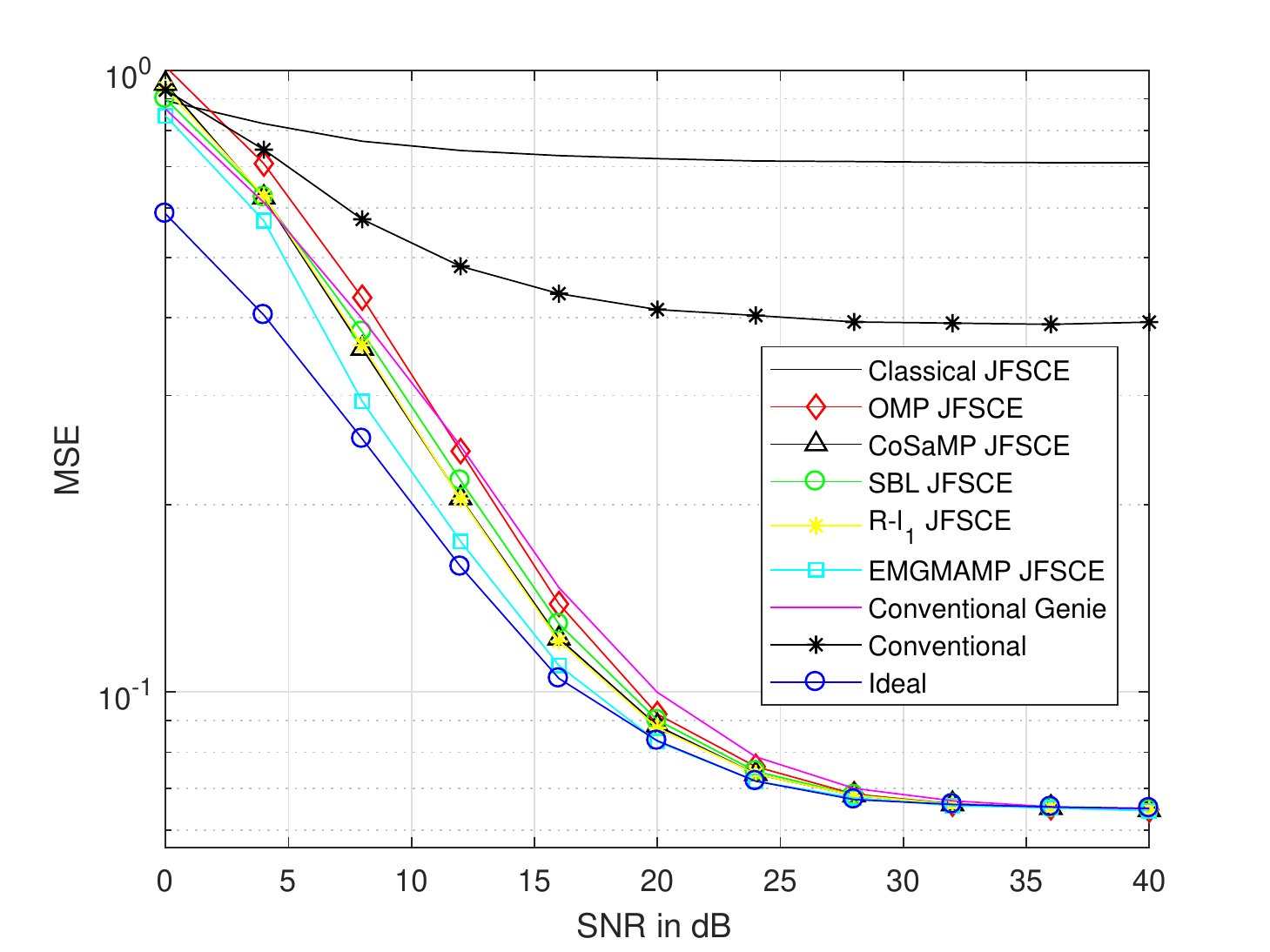}}
\caption{MSE versus SNR curves for different frame synchronization and channel estimation methods.}\label{MSE_Vs_SNR}
\end{figure}

Fig.~\ref{MSE_Vs_SNR} plots the mean-square error (MSE) after the equalization as defined in (\ref{Equalization_MSE}) for SSR-based JFSCE methods as well as the classical JFSCE method, conventional method and ideal method as a function of signal-to-noise ratio (SNR). As expected, the SSR-based JFSCE methods clearly outperform the classical JFSCE method and the conventional method by a large margin for most of the SNR range. Note that a common benchmark on the variance of any unbiased estimator of a deterministic parameter is the Cramer-Rao lower bound (CRLB). The CRLB has also been used to evaluate the performance of different channel estimation algorithms in the literature \cite{CRLB_MIMO,CRLB_Sparse,CRLB_Sparse_Eldar}. However, the sparse estimation problem that we have in (10) is different than the other channel estimation problems in the literature such as those in \cite{CRLB_MIMO,CRLB_Sparse,CRLB_Sparse_Eldar}, where synchronization and sparse channel estimation problems are coupled together. Therefore, deriving the CRLB for (10) is complicated, and to our best knowledge it is an open problem. In place of the CRLB, we use the ideal method as an alternative baseline against which practical algorithms are compared. This is analogous to some other work such as~\cite{Genie}, where a genie-aided channel estimator is used as a benchmark to assess the performance of the different channel estimators. In the ideal method, it is assumed that the equalizer has the perfect channel knowledge, i.e., the locations (support set) as well as the amplitudes of the nonzero representation elements. Thus, it gives the ideal performance curve on the MSE performance of different JFSCE and conventional methods discussed in the paper.

Note that the simulated channel in Fig.~\ref{channel:fig} has the strongest component at location 14 and it turns out that for the conventional method $\hat{\bar{D}}=514$ is obtained as the estimate of the actual frame boundary of $\bar{D}=500$. This causes the conventional method to perform poorly compared to the SSR-based methods. Also plotted in Fig.~\ref{MSE_Vs_SNR} is the genie-aided conventional method where the frame boundary is assumed to be known at the receiver. In this case the conventional method is capable of achieving performance close to SSR-based JFSCE methods.

For most of the SNR range, the EMGMAMP-based JFSCE method provides the best performance among the SSR-based JFSCE methods. At lower SNRs, EMGMAMP-based JFSCE method performs the best and is close to the ideal method followed by the R-$\ell_1$-based JFSCE method, SBL-based JFSCE method, CoSaMP-based JFSCE method and OMP-based JFSCE method. The superiority of EMGMAMP-based JFSCE method can be attributed to the fact that it near-optimally learns and exploits the signal priors. It is successful in learning a reasonable approximation of the unknown true probability density function from the noisy observations $\yb$ unlike other SSR-based JFSCE methods. For instance, the SBL-based JFSCE method needs apriori information. At higher SNR, all the SSR algorithms perform comparably. 

\begin{figure}[t!]
\centering
\centerline{\includegraphics[scale=0.8]{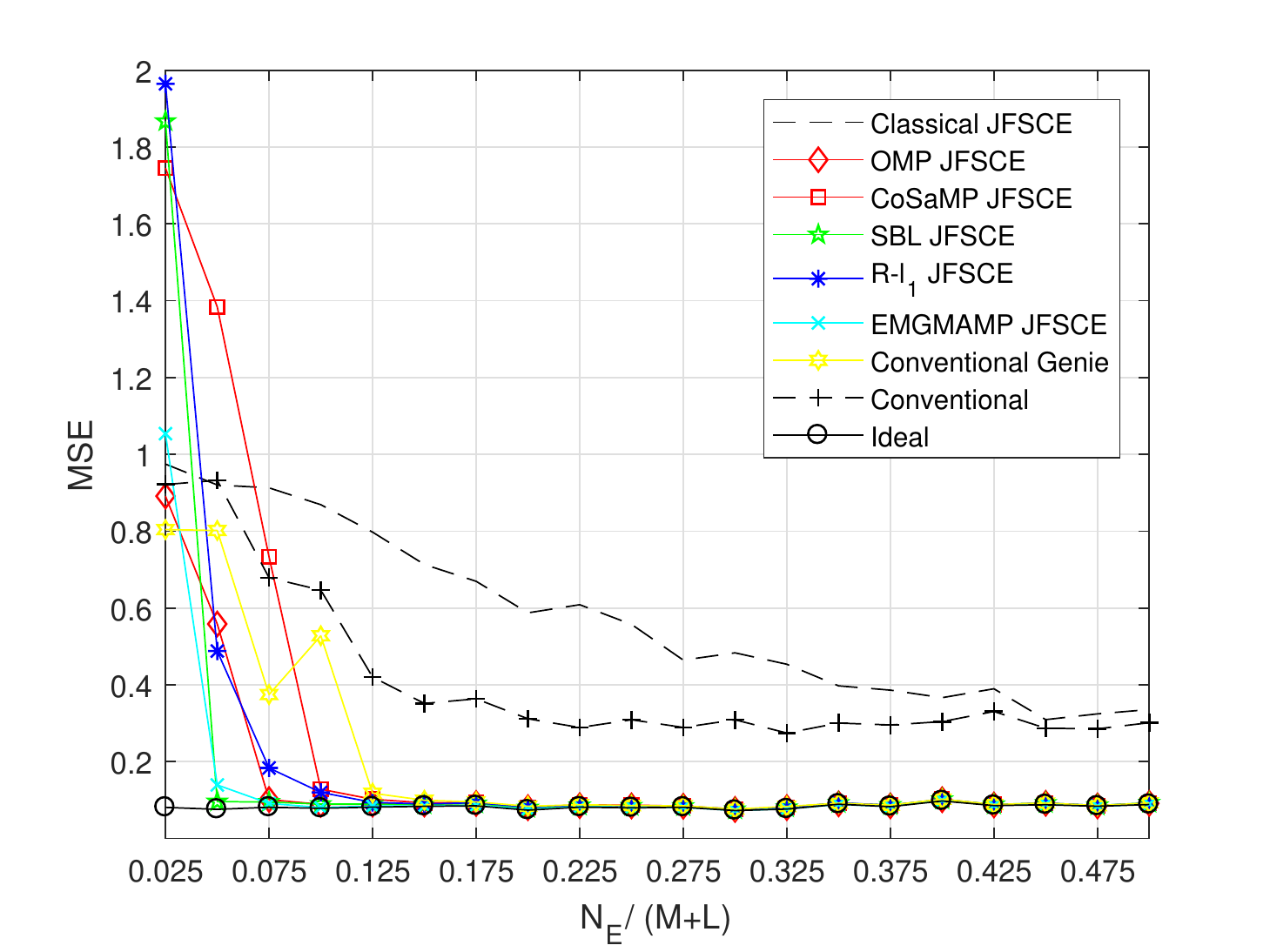}}
\caption{MSE versus the ratio of $N_E$ and $M+L$ curves for different frame synchronization and channel estimation methods. The SNR and length of $M+L$ is fixed to 20 dB and 1100, respectively.}\label{MSE_Vs_NE)}
\end{figure}

\subsection{Number of Measurements Required}
Fig. \ref{MSE_Vs_NE)} plots the MSE performance as a function of the number of measurements, $N_E$, with the length, $M+L$, of the vector $\tilde{\hb}$ fixed to 1100. A key result from the SSR literature states that, the sparse vector $\tilde{\hb}$ can be recovered with a high probability using variety of algorithms provided that the number of measurements $N_\text{E}$ satisfies the relation:
\begin{equation}\label{NE_Required}
N_E \geq c k \log\left(\frac{M+L}{k}\right)
\end{equation}
where c is a small constant and $k$ is the sparsity level of the vector $\hb$. As shown in Fig.~\ref{MSE_Vs_NE)}, the performances of all of the SSR-based JFSCE methods are identical when the number of measurements, $N_E$, is high. In fact, the performance of the classical JFSCE method tends towards the ideal performance if $N_E$ is greater than $M+L$, resulting in high overhead. This is where the SSR-based JFSCE methods triumph over the classical JFSCE method. The SSR-based JFSCE methods successfully recover even in an under-determined setting ($N_E << M+L$). With decreasing $N_E$, the MSE performance starts to decrease. Among the investigated SSR-based JFSCE methods, EMGMAMP-based JFSCE method performs well even when the number of measurements available is as low as 0.050*(M+L).

\subsection{Algorithm Execution Time}
\begin{figure}[t!]
\centering
\includegraphics[scale=0.8]{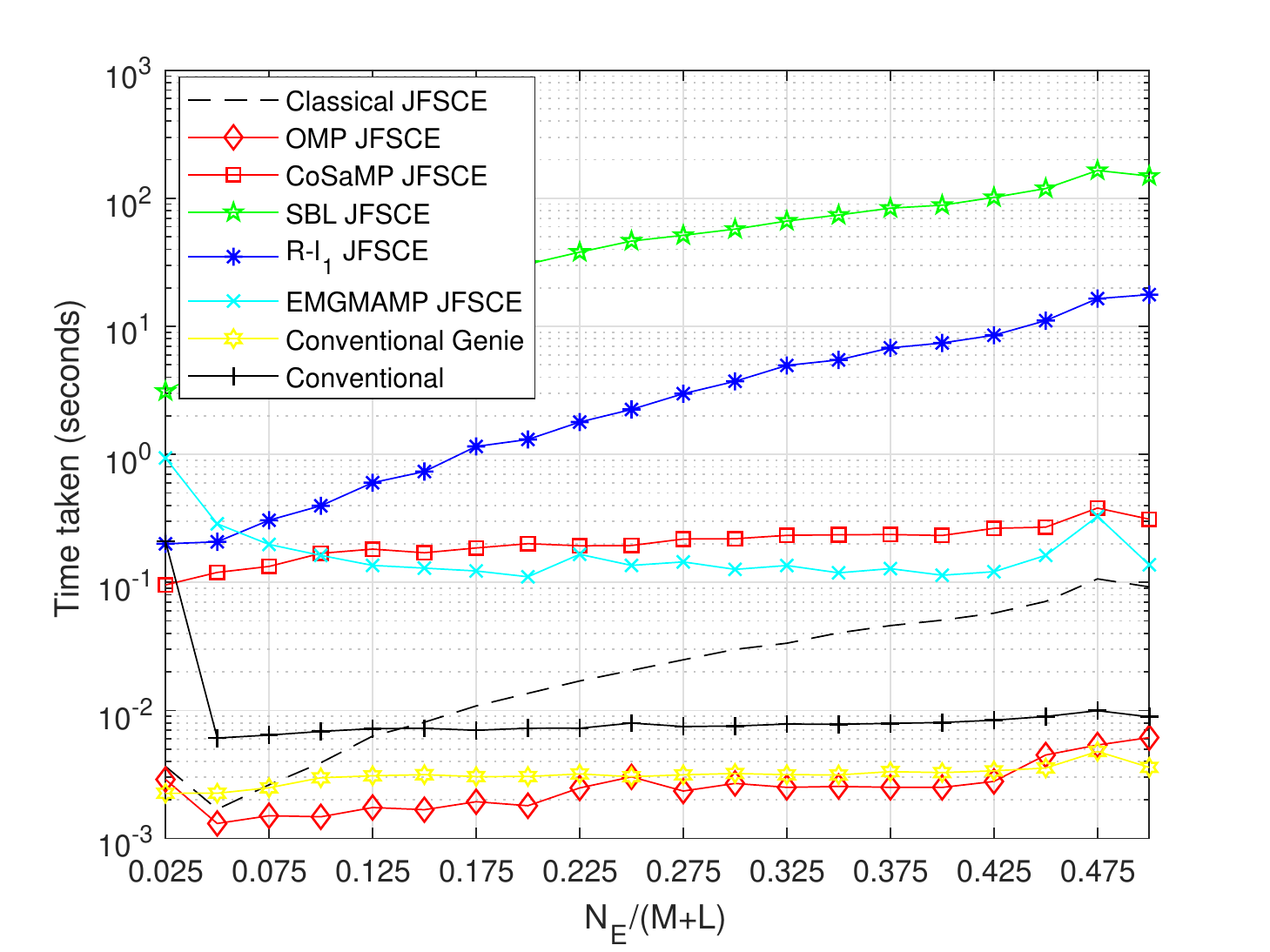}
\caption{Execution time versus the ratio of $N_E$ and $M+L$ curves for different frame synchronization and channel estimation methods. The SNR and length of $M+L$ is fixed to 20 dB and 1100, respectively.}\label{Time_Vs_NE}
\end{figure}

Fig. \ref{Time_Vs_NE} plots the time taken\footnote{Platform specifications: Intel(R) Core(TM) i7-4770 CPU $@$ 3.40 GHz, 12 GB RAM, x-64 based processor.} by the different frame synchronization and channel estimation methods as a function of the number of measurements, $N_E$, with $M+L$ being fixed. As evident, most of the methods scale linearly with the increase in $N_E$ except for the CoSaMP and OMP-based JFSCE methods. Based on these results, the CoSaMP and OMP-based JFSCE methods turn out to be the best in terms of execution time when compared to other methods. However, the CoSaMP and OMP-based JFSCE methods perform well only in the high SNR regime and when the ratio $N_E/(M+L)$ is high. Thus, the designer has to taken into account these trade-offs to decide on the choice of the SSR-based JFSCE method. 

\begin{figure}[t!]
\centering
\centerline{\includegraphics[scale=0.8]{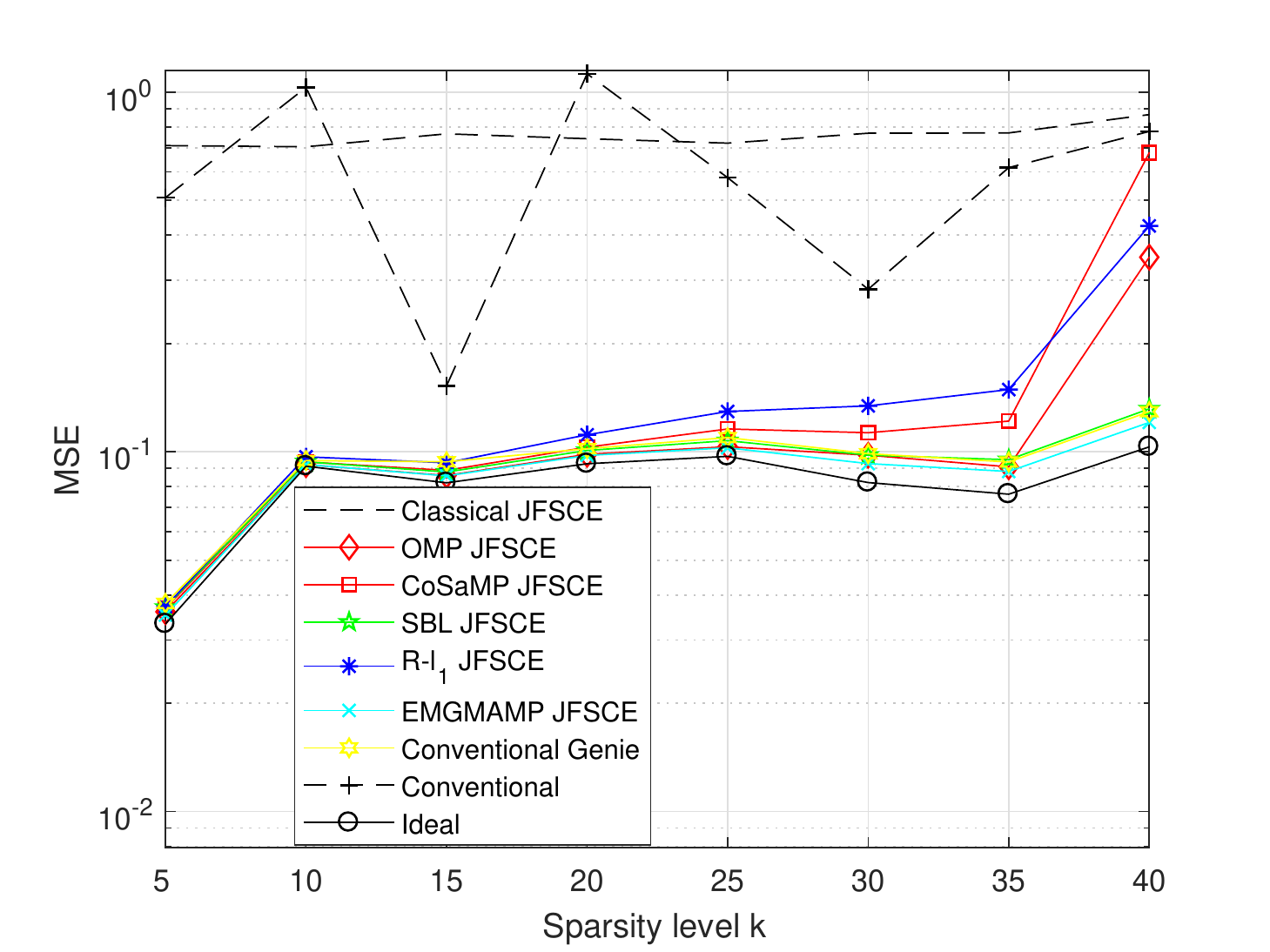}}
\caption{MSE versus sparsity level of $\tilde{\hb}$ curves for different frame synchronization and channel estimation methods. The other parameters are as follows: $N_E=148$, $M+L=1100$, and $SNR=20$ dB.}\label{MSE_vs_k}
\end{figure}

\subsection{Sparsity Level}
To illustrate the effect of the sparsity level with a fixed number of measurements, we performed the following experiment. We repeated the previous simulation with different sparsity levels, $k$, and fixed $N_E$ = 148 and $M+L$ = 1100. In other words, we simulated a random CIR, both the support set and amplitude being random, as opposed to a fixed CIR as shown in Fig. \ref{channel:fig}, by controlling the number of non-zero elements in $\hb$. Note that even if $\hb$ is not sparse, the extended vector $\tilde{\hb}$ will still remain sparse. Fig. \ref{MSE_vs_k} plots the MSE performance with the increase in the number of non-zero elements in $\hb$. For fixed $N_E$ and $M+L$, the MSE performance starts to deteriorate as the sparsity level of $\hb$ increases. This phenomenon can be related to (\ref{NE_Required}). The SSR-based JFSCE methods fail to recover the sparse channel vector $\tilde{\hb}$ as the sparsity level increases with fixed $N_E$. The MSE performance can be improved with increasing $N_E$. This implies that the designer needs to know an upper bound on the sparsity level. Knowing an upper bound on the sparsity level often helps to decide on the number of measurements required for successful recovery of the sparse channel vector. 
The reason why the MSE for the conventional method varies significantly is because the performance of the conventional method highly depends on the location of the highest peak of the CIR. If the highest peak is not at the start of the randomly generated CIR, then an error is made when determining the frame boundary. This leads to severe degradation in performance resulting in a non smooth curve as seen in Fig. 9. On the other hand, the genie-aided conventional method performs similar to the SSR-based JFSCE methods.

\subsection{Computational Complexities}
The computational complexities of different methods in terms of $N_\text{E}$, $(M+L)$ and sparsity level are shown in Table \ref{tab:Comp}. Note that $M << N_\text{E}$ and $k$ and $iter$ refer to the sparsity level and number of iterations the algorithm takes to converge, respectively. 


\begin{table}[t!]
\caption{Computational Complexity }
\begin{center}
\begin{tabular}{ |p{6.5cm}||p{4cm}|  }
 \hline
 Algorithm & Complexity\\
 \hline
 Classical JFSCE method & $\mathcal{O}(N_E^3)$ \\
 \hline
 OMP-based JFSCE method~\cite{OMP} & $\mathcal{O}(N_E\times(M+L)\times k)$\\
 \hline
 CoSaMP-based JFSCE method~\cite{cosamp} & $\mathcal{O}(N_E\times(M+L)\times iter)$ \\
 \hline
 R-$\ell_1$-based JFSCE method \cite{Rl1}  &  $\mathcal{O}(N_E^2\times(M+L)^3\times iter)$  \\
 \hline
 SBL-based JFSCE method~\cite{SBL}&   $\mathcal{O}((M+L)^3)$  \\
 \hline
 EMGMAMP-based JFSCE method~\cite{EMGMAMP} & $\mathcal{O}(N_E \times (M+L))$\\
 \hline
\end{tabular}\label{tab:Comp}
\end{center}
\end{table}

\section{Experimental Results with USRPs}

\begin{figure}[!tbp]
\centering
\includegraphics[scale=.2]{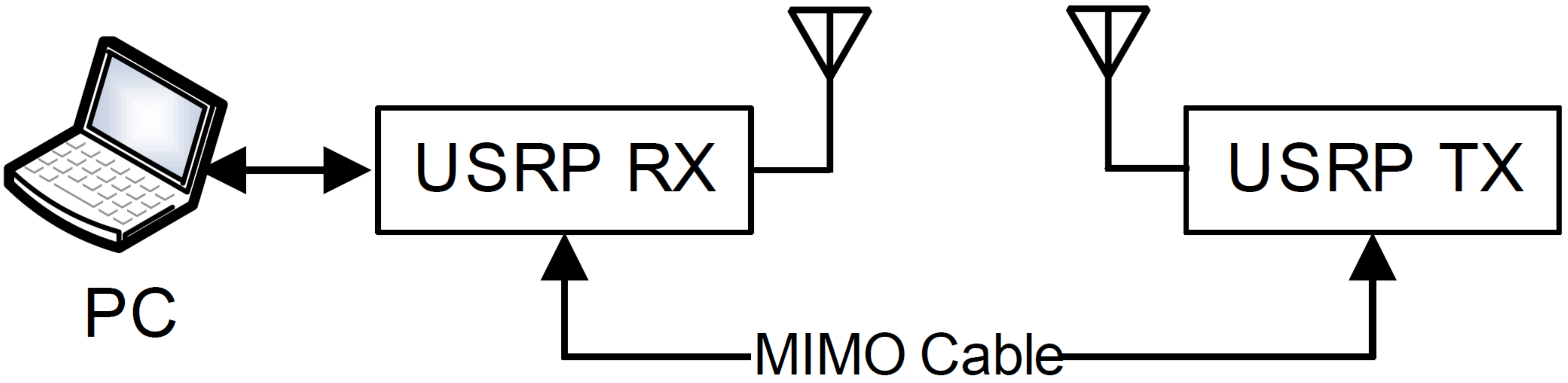}
\caption{System model for USRP experiments.}
\label{model:fig}
\end{figure}

The OMP-based JFSCE method as well as conventional method and classical JFSCE method were also implemented on our testbed based on USRP N210s~\cite{ettus} as shown in Fig.~\ref{model:fig}. The receiver USRP is connected to the PC with a gigabit ethernet connection. The transmitter USRP is connected to the receiver USRP with the multi-input multi-output (MIMO) cable. The digital baseband signal produced at the receiver USRP is sent to the PC by the gigabit ethernet connection. The digital baseband signal produced at the PC goes through the gigabit ethernet link and the MIMO cable to reach the transmitter USRP.

\subsection{Implementation of the Testbed Using Labview\texttrademark and MATLAB\texttrademark}
In~\cite{ozdemir&milcom:2017}, we have implemented both the baseband processing of the signals as well as the communication with the USRP devices in MATLAB\texttrademark based on QPSK transmitter~\cite{qpsk1} and receiver~\cite{qpsk2} examples. Here, we extend our work in~\cite{ozdemir&milcom:2017} so that Labview\texttrademark is used to communicate with the transmitter and receiver USRPs. At the transmitter, Labview\texttrademark's Matlab script is used to generate random data consisting of $P-1$ data frames and one training frame and the same data is transmitted over and over in infinite while loop. The data is  saved into the PC in MATLAB\texttrademark format so that the receiver can acquire the training frame to perform frame synchronization and channel estimation. This is possible because the same PC is used for both the transmitter and the receiver. The receiver also contains a while loop where the incoming data from the receiver USRP is processed in Matlab script and another Matlab script is placed outside the loop for initialization purposes. The code is available at~\cite{JFSCEcode} and it consists of one transmitter vi, one receiver vi and several MATLAB\texttrademark files that are called within the Matlab scripts mentioned above. 

\subsection{Transmitter for JFSCE Experiments}

The training frame period is selected to be $P=10$. Data frames contain $M=100$ whereas the training frame contains $\tilde{M}=147$ randomly generated QPSK symbols. Let us assume that $\bar{M}=(P-1)M+\tilde{M}$. Therefore, a total of $\bar{M}=1047$ QPSK symbols are generated at the transmitter. The symbols are up-sampled by a factor of 4 and passed through a root raised cosine transmit (RRC) filter. Thus, for every $\bar{M}$ QPSK symbols, $4\bar{M}$ samples are generated. 

Given the small bandwidth used in our experiments, we mostly observe a single-path channel where equalization is not needed. In order to test the proposed method under different multi-path environments, we manually insert a 2-tap symbol-spaced CIR denoted by $\hb^{(i)}[n]$ at the baseband transmitter after RRC filtering where
\be 
\label{hin:eq}
\delimiterfactor=1200 
\hb^{(i)}[n] = \left\{%
\begin{array}{ll}
1 &\textrm{if }n=0,\\
0.7 &\textrm{if }n=i,i=1,2,\cdots\\
0 &\textrm{otherwise}.
\end{array}%
\right.
\ee

\begin{figure}[!t]
\centering
\includegraphics[scale=1.2]{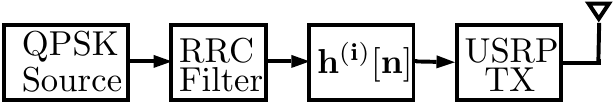}
\caption{Transmitter diagram used for JFSCE experiments.}
\label{modeltx:fig}
\end{figure}

To simplify the transmitter, the same $\bar{M}$ symbols are transmitted repeatedly. We note that although $\bar{M}$ QPSK symbols are repeated, because of RRC and $\hb^{(i)}[n] $ filtering, the first $4\bar{M}=4188$ samples will be different from the second $4\bar{M}$ samples generated which repeat afterwards. We discard the first $4\bar{M}$ samples and the second $4\bar{M}$ samples are transmitted repeatedly in the 900 MHz ISM band at a sampling rate of $f_{\mathrm{s}}=200$~kHz. A block diagram of the basic operations at the transmitter is shown in Fig.~\ref{modeltx:fig}.

\subsubsection{Receiver for JFSCE Experiments}
At the receiver side, the same sampling rate of 200 kHz is used as well. The automatic gain control (AGC) block is followed by the receiver RRC filter. The RRC filter reduces the the over-sampling factor from 4 down to 2. The RRC filter is followed by coarse frequency compensation where the fourth power of the signal is taken. Because the transmitted signal is QPSK modulated, there exists a peak at the FFT bin corresponding to 4 times the carrier frequency offset between the transmitter and the receiver. This offset is estimated and then compensated. If the FFT size is increased, the frequency resolution can be increased as well; however, there will be a residual frequency offset. The phased lock loop (PLL) based fine frequency compensation block~\cite[Chapter~7]{rice:book} following the coarse frequency compensation aims to compensate for this residual frequency offset. The fine frequency compensation block is followed by a timing recovery block which is also PLL based~\cite[Chapter~8]{rice:book}. During timing recovery, over-sampling is reduced to 1 and the receiver clock is synchronized to the transmitter clock. At this point, one sample corresponds to one symbol and the samples are buffered at the output of the timing recovery block and processed at $\bar{M}$ samples at a time. 

The channel estimation and equalization block is the last block where the methods described in this paper are implemented. This block accepts $\bar{M}$ samples at a time; however, it keeps the previous 2$\bar{M}$ samples in memory. Each time this block is executed, $\bar{M}$ samples at the center of the total  $3\bar{M}$ samples are processed. The first block is needed as the equalizer needs the previous samples as initial samples of the central $\bar{M}$ samples. The third block is needed due to the possibility that the training frame occurs towards the end of the central block and it extends into the third block.

Recall that in our frame structure, $\tilde{M}$ samples are periodically inserted after every $(P-1)M$ samples. Because the size of the training frame is different from the size of the data frame, the channel estimation and equalization block needs to keep track of the frame number and collect $\tilde{M}$ samples at the correct frame delay which was our assumption in (\ref{barD:eq}) that delay $mM$ is available to us and we are only unaware of $\bar{D}$ where $0 \leq \bar{D} < M$. To satisfy this assumption in practice, we correlate the training frame with the received samples and determine the sample where the training frame starts and identify the integer $m$ in (\ref{barD:eq}) from this sample.\footnote{The same correlation is used for frame synchronization in the conventional method. However, the conventional method needs to do it more often than the proposed method. For the proposed method, this correlation needs to be performed every time $m$ is expected to change. On the other hand, for conventional methods this correlation is required to be performed every time the delay $D$ is expected to change.} Without loss of generality, we assume that the sample delay $0\leq D < \bar{M}$ is the delay associated with the training frame and $0 \leq m \leq P$. As an example if $0 \leq D < M $,  then the first $\tilde{M}$ samples of the central $\bar{M}$ samples correspond to the training frame to be used for the proposed method. Table~\ref{tab:1} lists all the possible values of $D$ and corresponding training frame samples to be used for channel estimation and frame synchronization.\footnote{A word of caution, the training frame samples that are listed in here still contain the unknown delay $\bar{D}$.} Note that, when $PM \leq D < \bar{M} $ where the delay is towards the end of $\bar{M}$ samples, the training frame samples extend from the central $\bar{M}$ samples into the third $\bar{M}$ samples being buffered in the channel estimation and equalization block.

\begin{table}[h!]
    \begin{center}
        \begin{tabular}{ccc}\hline
        Sample Delay   & Frame Delay & Training Frame \\
        $D$   & $m$ & Samples \\
        \hline         
\noalign{\vskip 1mm}          $0 \leq D < M $ & 0 & $[1,\tilde{M}]$ \\
\noalign{\vskip 1mm}          $M \leq D < 2M $ & 1 & $[M+1,M+\tilde{M}]$ \\
\noalign{\vskip 1mm}          $\vdots$ & $\vdots$ & $\vdots$ \\
\noalign{\vskip 1mm}          $(P-1)M \leq D < PM $ & P-1 & $[(P-1)M+1,\bar{M}]$ \\
\noalign{\vskip 1mm}          $PM \leq D < \bar{M} $ & P & $[PM+1,PM+\tilde{M}]$ \\
\hline  \noalign{\vskip 1mm} 
        \end{tabular}
        \caption{Training frame samples to be used for channel estimation and frame synchronization for different values of delay values}\label{tab:1}
    \end{center}
\end{table}\vspace{-1cm}

\begin{figure}[!t]
\centering
\includegraphics[scale=1.1]{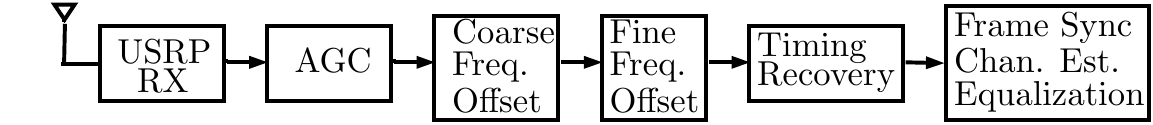}
\caption{Receiver diagram used for JFSCE experiments.}
\label{modelrx:fig}
\end{figure}

The training frame samples listed in Table~\ref{tab:1} are used to construct $\yb$ in (\ref{sys:eq}) and solve for the combined channel vector $\tilde{\hb}$. A block diagram of the basic operations at the receiver is shown in Fig.~\ref{modelrx:fig}.

\subsection{SDR Experimental Results on JFSCE}

The parameters used throughout this section are as follows: The channel estimates for the joint method are obtained using $N_{\mathrm{E}}=43$ measurements. The number of non-zero taps in the channel is upper bounded by $L+1=6$. The equalizer has 200 total number taps. 

\begin{figure*}[!t]
\centering
\includegraphics[width=5.5 in]{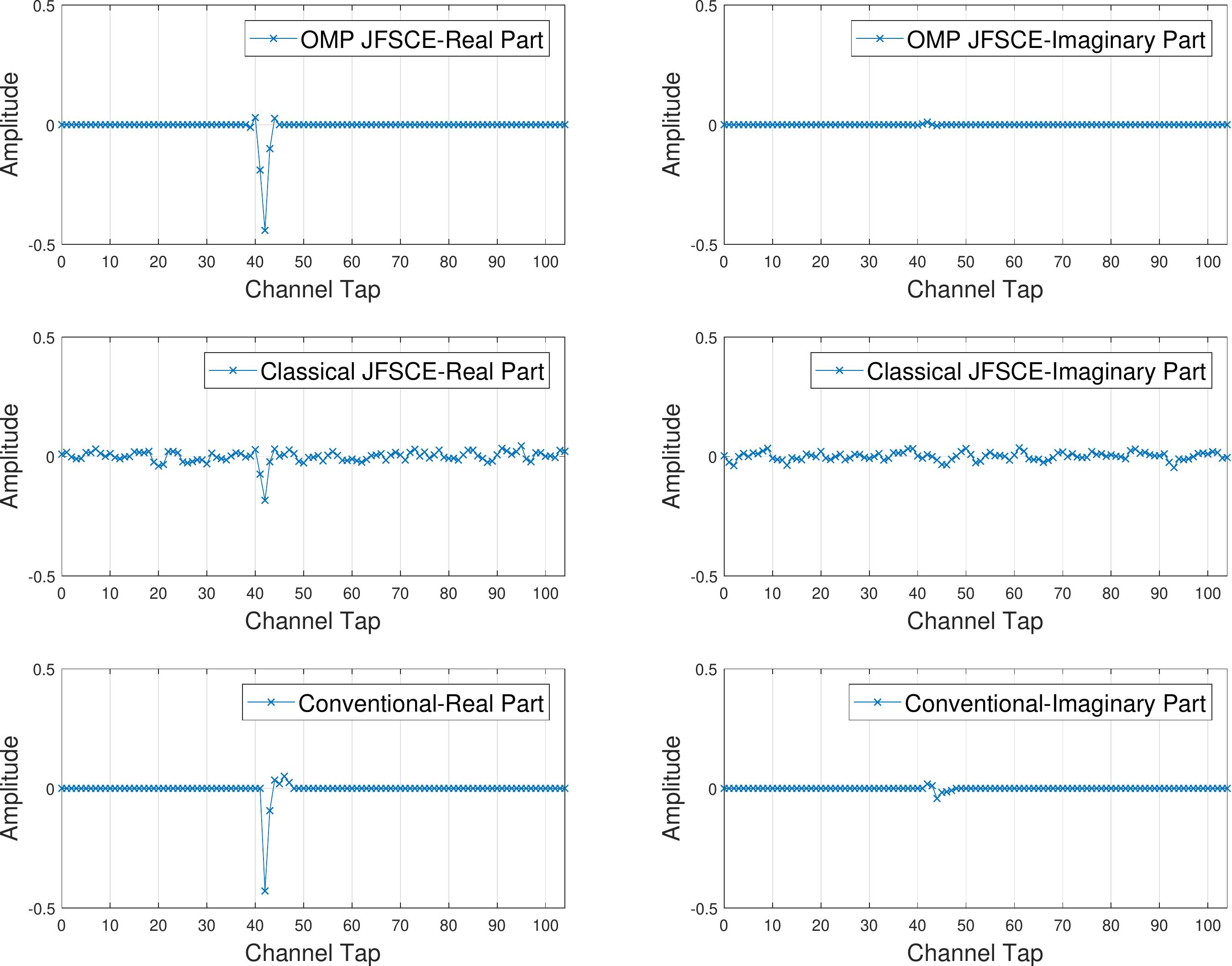}
\caption{The channel estimates obtained for $\hb^{(1)}[n]$ using OMP-based JFSCE method, classical JFSCE method, and conventional method.}
\label{channelestimates2:fig}
\end{figure*}

In Fig.~\ref{channelestimates2:fig}, the real and imaginary parts of the CIR estimates are shown separately where $\hb^{(1)}[n]$ as defined in (\ref{hin:eq}) is manually inserted at the transmitter side. The CIR estimates are obtained using the OMP-based JFSCE method, classical JFSCE method and the conventional method where the frame synchronization and channel estimation are performed separately. Both the OMP-based JFSCE method and the conventional method have 6 non-zero taps. We note that all three methods can accurately locate the frame boundary. The conventional method is allowed to contain non-zero taps only after the strongest tap. On the other hand, the OMP-based JFSCE method has more flexibility and contains non-zero taps on both sides of the strongest tap. The classical JFSCE method results in a noisy estimate of the channel where the channel taps spread all across the frame.

\begin{figure}[!htbp]
\centering
\includegraphics[scale =0.8]{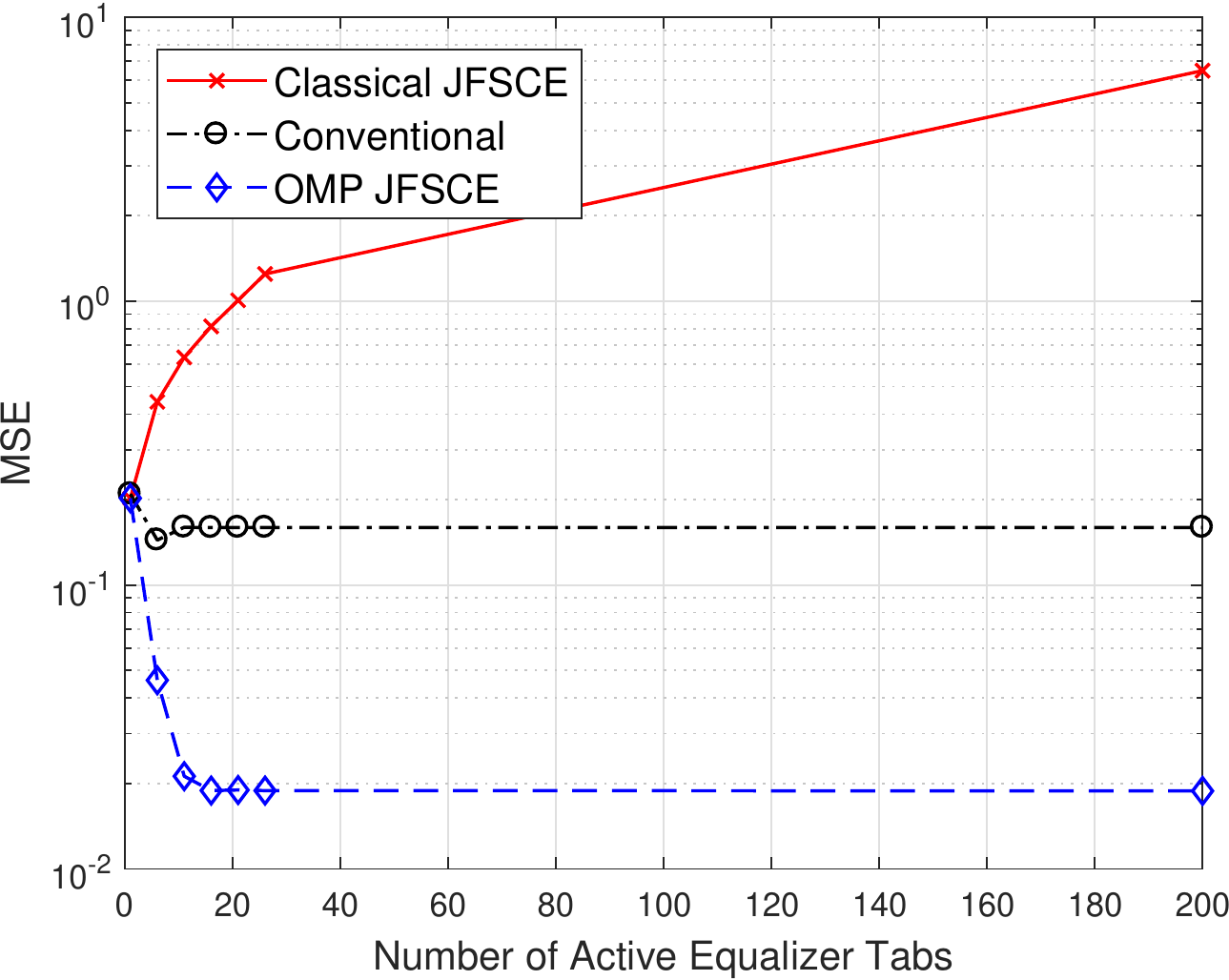}
\caption{The number of active equalizer taps versus MSE for different methods.}
\label{sim2:fig}
\end{figure}

To assess the performance of each method, we perform the sparse FIR linear equalization of~\cite{abbasi&globalsip:2015} as explained in Section~\ref{equalizer:sec}. The channel estimates from the three methods mentioned above are used to design the equalizer. The equalization is applied on all of the central $\bar{M}$ samples. Fig.~\ref{sim2:fig} shows the MSE results as a function of the number of active taps used in the equalizer out of a total of 200 taps. Note that the OMP-based JFSCE method outperforms the conventional method and the classical JFSCE method. For an accurate channel estimate, it is expected that the MSE will decrease as the number of active taps increase. When the channel estimate is not accurate as in the classical JFSCE method, the MSE may actually increase as the number of active taps increases. We also note that for the OMP-based JFSCE method, around 10 active taps are sufficient for the equalizer to converge to the optimum performance.

\begin{figure}[!tbp]
\centering
\includegraphics[scale=0.5]{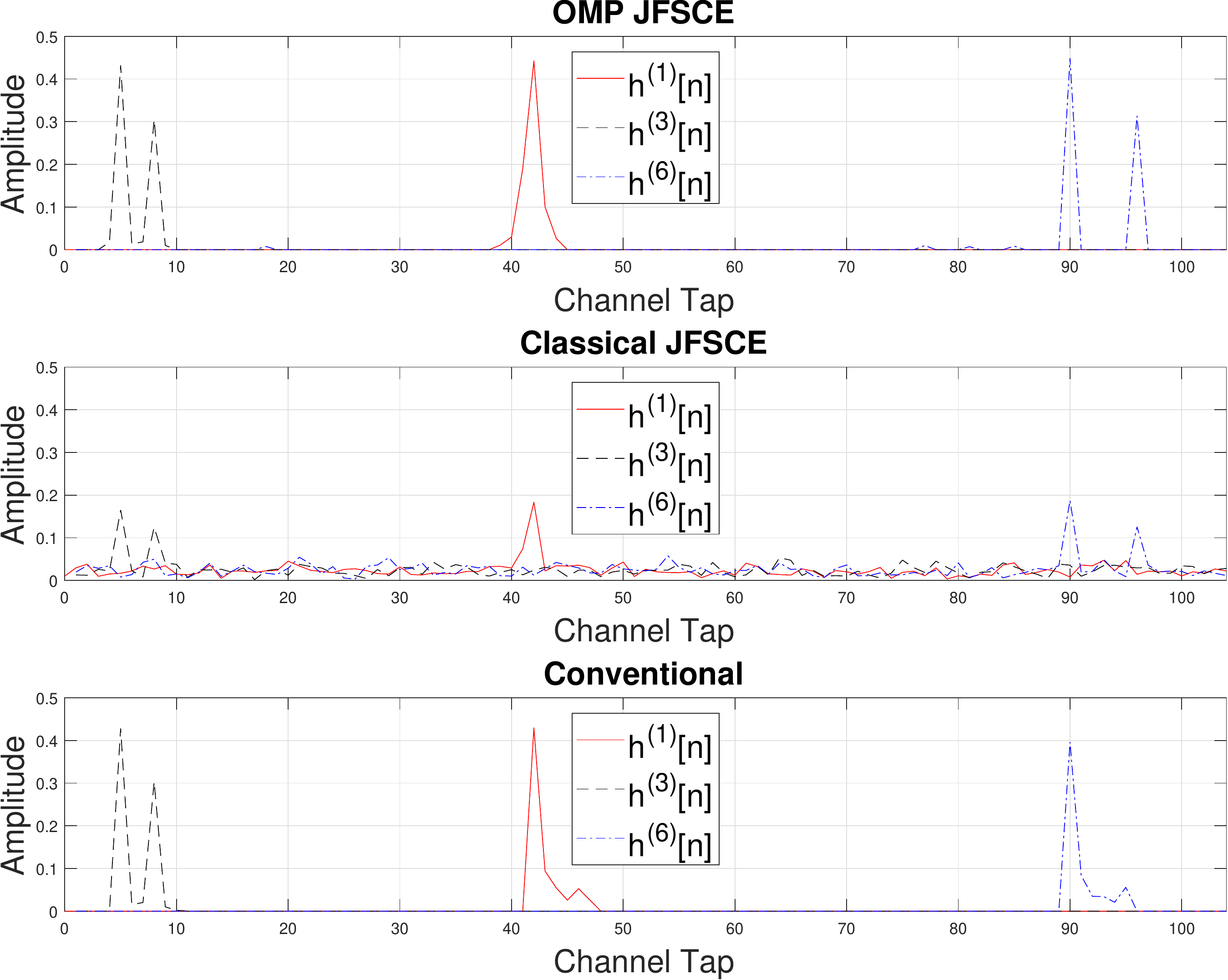}
\caption{The channel estimates obtained for $\hb^{(1)}[n]$, $\hb^{(3)}[n]$, and $\hb^{(6)}[n]$ using OMP-based JFSCE method, classical JFSCE method, and conventional method.}
\label{simic:fig}
\end{figure}

The CIR estimates in Fig.~\ref{simic:fig} contain results for both the manually inserted channel $\hb^{(1)}[n]$ as in Fig.~\ref{channelestimates2:fig} in addition to $\hb^{(3)}[n]$ and $\hb^{(6)}[n]$. When the manually inserted channel is updated, the Labview\texttrademark code is rerun and the frame boundary moves to a random location. In this figure, the absolute values of the CIR estimates are shown.  Our conclusions from Fig.~\ref{channelestimates2:fig} are the same for the results of $\hb^{(1)}[n]$. For $\hb^{(3)}[n]$, we observe that the OMP-based JFSCE method and the conventional method compute similar estimates. The OMP-based JFSCE method is allowed to have non-zero taps at any location. However, the conventional method is allowed to have 6 consecutive non-zero taps beginning with the strongest tap. When the manually inserted channel is $\hb^{(6)}[n]$, the estimate from the conventional method has only one strong path and the second tap is outside of the range that the method can handle. The OMP-based JFSCE method can handle this situation without any issues. 

\begin{figure}[!htbp]
\centering
\includegraphics[scale = 0.8]{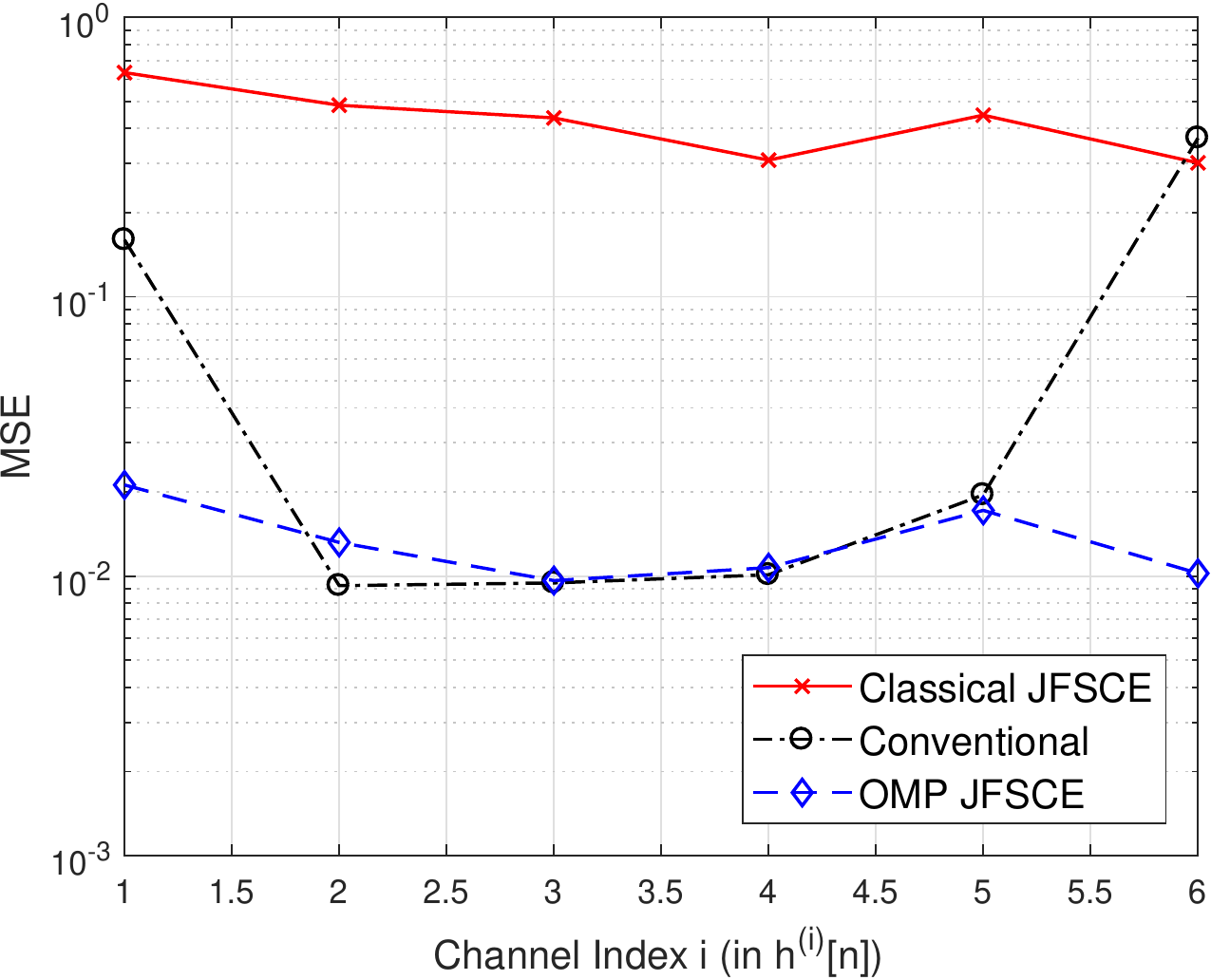}
\caption{MSE performance of different methods for different multipath environments.}
\label{simim:fig}
\end{figure}

Fig.~\ref{simim:fig} illustrates MSE results when the equalizer has 11 active taps and the index $i$ of the manually inserted channel $\hb^{(i)}[n]$ is varied. The OMP-based JFSCE method and the classical JFSCE method are slightly affected as the second tap of the channel is separated further away from the strongest tap. The conventional method has similar performance compared with the OMP-based JFSCE method for $2 \leq i \leq 5$. When $i=5$, the conventional method can still estimate the second tap as it is allowed to have 6 non-zero taps. When $i>5$, we would expect that its performance would decrease as suggested by Fig.~\ref{simic:fig}. The OMP-based JFSCE method performs well both inside and outside of the range $2 \leq i \leq 5$. As expected, the classical JFSCE method performs worst similar to the previous results.

\section{Conclusion}

In this paper, we proposed a JFSCE framework for sparsity-aware joint frame synchronization and channel estimation. Our simulation results and experimental USRP results demonstrated that proposed SSR-based JFSCE methods achieve superior performance compared to the conventional method where frame synchronization and channel estimation are performed separately. In the proposed JFSCE method, training frames are inserted between data frames which increases the overhead; however, this overhead can be minimized by increasing the period of the training frame by taking the coherence time of the channel into consideration. Future research includes extensions to multi-antenna and multi-user systems.

\bibliographystyle{IEEEtran}
\bibliography{draft}

\end{document}